\documentclass[aps,prd,twocolumn,superscriptaddress,showpacs]{revtex4}
\usepackage{graphicx}
\usepackage{lscape}
\usepackage{epsfig,epsf}
\usepackage{amsmath}
\usepackage{amsthm}
\usepackage{amsfonts}
\usepackage{amssymb}
\usepackage{dsfont}
\usepackage{multirow}
\usepackage{appendix}
\usepackage{slashed}
\usepackage[active]{srcltx}
\usepackage{psfrag}
\usepackage{multirow}

\newcommand{\be}{\begin{equation}}
\newcommand{\ee}{\end{equation}}
\newcommand{\bea}{\begin{eqnarray}}
\newcommand{\eea}{\end{eqnarray}}

\def\R1{\varepsilon_1}
\def\E8{\varepsilon_8}

\def\s1{\hat s}

\newcommand{\bd}{\begin{displaymath}}
\newcommand{\ed}{\end{displaymath}}

\def\R1{\varepsilon_1}
\def\E8{\varepsilon_8}

\def\beq{\begin{equation}}
\def\eeq{\end{equation}}
\def\bea{\begin{eqnarray}}
\def\eea{\end{eqnarray}}
\def\beeq{\begin{eqnarray}}
\def\eeeq{\end{eqnarray}}

\def\nnb{\nonumber}

\def\rar{\rightarrow}
\def\nnb{\nonumber}

\def\ba{\begin{array}}
\def\ea{\end{array}}

\def\xis0{{\Xi^{*0}}}

\def\g5{\gamma_5}

\begin{document}

\title{ Effects of vector leptoquarks  on  $ \Lambda_b \rightarrow \Lambda_c~\ell ~ \overline{\nu}_\ell$ decay }
\date{\today}
\author{K.~Azizi}
\affiliation{Department of Physics, University of Tehran, North Karegar Avenue, Tehran
14395-547, Iran}
\affiliation{Department of Physics, Do\v{g}u\c{s} University, Ac{\i}badem-Kad{\i}k\"{o}y, 34722
Istanbul, Turkey}
\author{A.~T.~Olgun}
\affiliation{Vocational School, Tuzla Campus, Istanbul Okan University,Tuzla, 34959 Istanbul ,Turkey}
\author{Z.~Tavuko\u glu}
\affiliation{Vocational School, Tuzla Campus, Istanbul Okan University,Tuzla, 34959 Istanbul ,Turkey}

\begin{abstract}
Experimental data on  $ R(D^{(*)}) $, $ R(K^{(*)}) $ and $ R(J/\psi) $, provided by different collaborations, show sizable deviations from the SM predictions.  To describe these anomalies many new physics scenarios have been proposed. One of them is leptoquark model with introducing the vector and scalar leptoquarks coupling simultaneously to the quarks and leptons. To look for similar possible anomalies in baryonic sector, we investigate the effects of a vector  leptoquark $U_3 (3,3, \frac{2}{3})$ on various physical quantities related to the   tree-level  $ \Lambda_b \rightarrow \Lambda_c \ell ~ \overline{\nu}_\ell$ decays ($ \ell=\mu, ~\tau $), which proceed via $ b \rightarrow c~\ell ~ \overline{\nu}_\ell$ transitions at quark level. We calculate the differential branching ratio, forward-backward asymmetry  and longitudinal polarizations of lepton  and $\Lambda_{c}$ baryon at $ \mu $ and $ \tau $ lepton channels in  leptoquark model and compare their behavior with respect to $ q^2 $  with the predictions of the standard model (SM). In the calculations we use the form factors calculated in full QCD as the main inputs and take into account all the errors coming from the form factors and model parameters. It is observed that, at $ \tau $ channel,  the $ R_A $ fit solution to data related to  the leptoquark model sweeps some regions out of the SM band but it has a considerable intersection with the SM predictions. The  $ R_B$ type solution gives roughly the same results with the those of the SM on $ DBR(q^2)-q^2$. At $ \mu $ channel, the leptoquark model gives consistent results  with the SM predictions and existing experimental data on the behavior of $ DBR(q^2)$ with respect to $ q^2 $. As far as the $ q^2 $ behavior of the $ A_{FB}(q^2) $ is concerned, the two types of  fits in leptoquark model  for $ \tau $ and the  predictions of this model at  $ \mu $ channel  give  exactly the same results as the SM.  We also investigate the behavior of  the  parameter $ R(q^2)  $ with respect to $ q^2 $ and the value of  $ R(\Lambda_c)  $  both in vector leptoquark and SM models. Both types  fit solutions lead to  results that deviate considerably   from the SM predictions on $R(q^2)- q^2 $ as well as $ R(\Lambda_c)  $.    Future experimental data on $R(q^2)- q^2 $ as well as $ R(\Lambda_c) $, which would be available after measurements on  $ \Lambda_b \rightarrow \Lambda_c \tau ~ \overline{\nu}_\tau$ channel,  will be very helpful. Any  experimental deviations from the SM predictions in this channel  will strengthen the importance of the tree-level hadronic weak transitions as good probes of the new physics effects beyond the SM (BSM). 
\end{abstract}


\maketitle

\section{Introduction}
\label{Sec:Int}
The search for new physics (NP) effects BSM constitutes one of the main directions of the research in particle physics. The direct search for NP effects and the predicted new particles have received null results so far and these effects have been excluded up to  a few TeV.  However, recently, there were recorded  significant deviations of the experimental data on some parameters of the weak decays of some hadrons from the SM predictions. These deviations may be considered as signs for the NP effects and are in agenda of many experimental and theoretical groups. Hence,  the weak and semileptonic hadronic decays are receiving special attention. Among these decays are the 
semileptonic mesonic  $ B \rightarrow D^{(*)} \ell \overline{\nu}_{\ell}$  and $ B_c \rightarrow J/ \psi (\eta_c)  \ell \overline{\nu}_{\ell}$ tree-level decays as well as the loop-level $ B \rightarrow K^{(*)} \ell^{+} \ell^{-} $  transitions. These channels provide major contributions to both  re-test the SM and  investigate the  NP effects.  In the SM, these decays occur by couplings to $W^{\pm}$, $ Z $ and $ \gamma $ which are assumed to be universal for all leptons. 
Normally,  different masses of the charged leptons lead to  different results in the branching fractions of the semileptonic decays including these leptons.  Extra discrepancies with the  SM  predictions on parameters of these decays suggest  lepton flavor universality violation (LFUV), which may be considered as the presence of the new particles BSM.
In particular the $\tau$ channel, because of the  larger mass of $\tau$,  is highly sensitive to the  contributions of hypothetical new particles like  charged Higgs boson, which appear in  leptoquark (LQ) or other NP models.

Over the past two decades, the experimental measurements on different parameters  related to the aforementioned decay channels have greatly improved at B factories. 
The branching ratio of  $B \rightarrow D^{(*)}\ell^{-} \overline{\nu}_{\ell}$ decay, which is very sensitive to the  NP scenarios,
 is considered as one of the major sources of the LFUV. The  parameters $ {\cal R }(D) $ and $ {\cal R }(D^{(*)}) $, defined as 
\begin{equation}
R(D^{(*)})= \frac{{ \cal B}(B\rightarrow D^{(*)} \tau \overline{\nu}_{\tau})}{{\cal B}(B\rightarrow D^{(*)} e(\mu)\overline{\nu}_{e(\mu)})},
\end{equation}
and with the  average values measured by BaBar, Belle and LHCb Collaborations \cite{Amhis:2019ckw}
\begin{eqnarray} 
R(D) = 0.340\pm 0.027\pm 0.013,
 \end{eqnarray}
 and
 \begin{eqnarray} 
R(D^{*})=0.295\pm 0.011\pm 0.008,
\end{eqnarray} 
indicate respectively  $1.4\sigma$ and $2.5\sigma$  deviations from the related SM predictions.
Another source is 
\begin{eqnarray}
R_{K^{(*)}}\equiv\frac{ BR(B\rightarrow K^{(*)}\mu^+\mu^-) }{BR(B\rightarrow K^{(*)}e^+e^-)}. 
\end{eqnarray}
The LHCb collaboration measured 
\begin{eqnarray}
R_{K}=0.745^{+0.090}_{-0.074}(\mbox{stat})\pm 0.036(\mbox{syst}), 
\end{eqnarray}
in the interval  $q^2\epsilon[1,6]$~GeV$^2$~\cite{Aaij:2014ora}, 
\begin{eqnarray}
R_{K^{*}}=0.66^{+0.11}_{-0.07}(\mbox{stat})\pm 0.03(\mbox{syst}),
\end{eqnarray}
 in the region  $q^2\epsilon[0.045,1.1]$~GeV$^2$ and 
\begin{eqnarray}
R_{K^{*}}=0.69^{+0.11}_{-0.06}(\mbox{stat})\pm 0.05(\mbox{syst}),
\end{eqnarray}
for $q^2\epsilon[1.1,6]$~GeV$^2$~\cite{Aaij:2017vbb}, indicating the deviations from SM expectations  by $(2.2-2.6) \sigma$ ~\cite{Bordone:2016gaq,Descotes-Genon:2015uva}.
The recent LHCb data on $  R (J/ \psi)$ for decay of $B_c \rightarrow J/ \psi \ell\overline{\nu}_{\ell}$ \cite{Aaij:2017tyk}:
\begin{equation}
  R (J/ \psi)= 0.71 \pm 0.17(stat) \pm 0.18(syst)
\end{equation}
shows serious deviations from the SM predictions \cite{Aaij:2017tyk,Cohen:2018dgz,Rui:2016opu,Dutta:2017xmj,Leljak:2019eyw,Murphy:2018sqg,Wen-Fei:2013uea}.  A recent more precise SM prediction made in \cite{Azizi:2019aaf}, $   R (J/ \psi)=0.25\pm 0.01 $, supports the existing tension between the SM theory prediction and the experimental data. In this study, authors calculated $R(\eta_c)$  in $B_c \rightarrow J/ \eta_c \ell\overline{\nu}_{\ell}$ as well, which may be in agenda of different experiments in near future. Any deviations of the measured results from the SM predictions will increase the importance of the tree-level charged weak decays as possible probes of the NP effects (for further related studies see \cite{Issadykov:2018myx,Huang:2007kb,Kiselev:2002vz,Ivanov:2006ni,Wang:2008xt,Hu:2019qcn,Huang:2018nnq,Berns:2018vpl,Blanke:2019qrx}).

Experiments have mainly focused on the tree-level mesonic transitions based on $ b \rightarrow c~\ell ~ \overline{\nu}_\ell$, whilst similar discrepancies may be detected at tree-level baryonic transitions proceed via $ b \rightarrow c~\ell ~ \overline{\nu}_\ell$.  The semileptonic $ \Lambda_b \rightarrow \Lambda_c \ell ~ \overline{\nu}_\ell$ channel is one of the important ones that is expected to be in focus of much attention both experimentally and theoretically. The form factors of this transition as the main inputs to theoretically analyze this mode in SM and BSM are available using some methods and approaches. In Ref. \cite{Azizi:2018axf}, as an example, the related form factors were calculated in full QCD. Using these form factors, $R(\Lambda_c)=\frac{{ \cal B}(\Lambda_b\rightarrow \Lambda_c \tau \overline{\nu}_{\tau})}{{\cal B}(\Lambda_b\rightarrow \Lambda_c \mu\overline{\nu}_{\mu})}=0.31\pm 0.11$ was obtained, which needs to be checked in the experiment. 

Many models of new physics have been proposed to explain   the above mentioned experiment-SM anomalies. One of the most popular and on the agenda  new physics models that can play an important role  to solve these anomalies is the LQ model \cite{Buchmuller:1986zs,Dorsner:2016wpm}. LQs that naturally appear in several new physics model such as extended technicolor model \cite{Schrempp:1984nj}, compositeness \cite{compositenes}, Pati-Salam model \cite{Pati:1974yy}, grand unification theories with $SU(5)$ \cite{Georgi:1974sy} and $SO(10)$ \cite{Georgi:1974my} are hypothetical color-triplet bosons. LQs can carry both lepton (L) and baryon (B) quantum numbers with electric and color charges. These particles couple to both the leptons and quarks simultaneously and, as a result,  modify the amplitudes of the transitions that they are contributed. According to their properties under the Lorentz transformations, they are divided into two main categories: The spin 0  scalar leptoquarks  as well as the  spin 1  vector leptoquarks.
In this study, we  consider  a single vector leptoquark $U_3 (3,3, \frac{2}{3})$  which can provide a simultaneous explanation to  the  anomalies  in $b \rightarrow c$ and $b \rightarrow s$ transitions. The numbers inside the  bracket represent the SM gauge
group $ SU(3) \times SU(2)\times U(1) $ transformation properties: They refer to the color, weak and hyper-charge  representations, respectively. The vector leptoquarks were theoretically studied in \cite{Angelescu:2018tyl,Zhang:2019jax,Fajfer:2015ycq,Li:2016pdv,Zhang:2019jwp,Sahoo:2018ffv,Calibbi:2017qbu,Sahoo:2016pet,Assad:2017iib,Blanke:2018sro}.
Using the vector LQ,  $U_3 (3,3, \frac{2}{3})$, we calculate several observables such as  the differential branching ratio,
 the lepton forward-backward asymmetry, longitudinal polarization of lepton  and $\Lambda_{c}$ baryon  and ratio of differential branching ratios in $ \tau $ and $ \mu (e) $ channels, $R(\Lambda_c )$,  for the $ \Lambda_b \rightarrow \Lambda_c \ell ~ \overline{\nu}_\ell$ transition. 
 Using the form factors calculated in full theory, we numerically analyze the physical quantities both in SM and  vector LQ model and compare the obtained results with each other. Any future experimental data and their comparisons with the predictions of the present study will help us  check whether there exists any discrepancy with the SM predictions in the channel under question or not and, if this is the case, whether the anomalies can be described by the vector LQs or not?
 Note that, in Ref. \cite{Li:2016pdv}, a similar analysis on the tree-level  $ \Lambda_b \rightarrow \Lambda_c \tau~ \overline{\nu}_\tau$ decay  is done both in scalar and vector leptoquark scenarios using  the form factors calculated from the QCD sum rules  in HQET limit and  lattice QCD  with 2 + 1 dynamical flavours. 
Although there are studies on the polarization of the parent baryon $\Lambda_b$ as an observable in Refs. \cite{Kadeer,Bialas},   we neglect to discuss it  since it has been measured  by LHCb setup to be negligibly small \cite{Aaij:2013oxa}.

The outline of the paper is  as follows. In next section, we present the effective Hamiltonian responsible 
for the transitions under consideration both in the standard and LQ models. 
In section III, we depict the transition amplitude and matrix elements defining the  transition under study. 
In section IV, we calculate some physical quantities related to the baryonic  $ \Lambda_b \rightarrow \Lambda_c \ell ~ \overline{\nu}_\ell$ channel and numerically analyze  the results obtained. We compare the LQ model  predictions with those of the SM in this section. We reserve the last section for the summary and conclusions.


\section{The Effective Hamiltonian}
The hadronic transition of  $\Lambda_b \rar \Lambda_c \ell \overline{\nu}_{\ell} $ proceeds via $ b \rar c \ell \overline{\nu}_{\ell} $ at tree-level. The low-energy effective Hamiltonian defining this transition in  SM can be written as 
\begin{eqnarray} \label{HeffSM} 
{\cal H}^{eff}_{SM} &=& {G_F  
 \over \sqrt{2} } V_{cb} \bar{c}\gamma_\mu (1-\gamma_5) b \, \bar{\ell} \gamma^\mu (1-\gamma_5) \nu_{\ell}, 
\end{eqnarray}
where $G_{F}$ is the Fermi weak coupling constant and $V_{cb}$ is  one of the elements of the Cabibbo-Kobayashi-Maskawa (CKM) matrix.  Considering the LQ contributions of the exchange of vector multiplet $U_{3}^{\mu}$ at tree level, the effective Hamiltonian including the SM contributions and LQ corrections can be written as \cite{Zhang:2019jax,Fajfer:2015ycq}
\begin{eqnarray} \label{HeffSMLQ} 
{\cal H}^{eff}_{SM+LQ} &=& { G_F  V_{cb}
 \over \sqrt{2} }  \Big[C_{V} [\bar{\ell}\gamma_\mu (1-\gamma_5) \nu_{\ell}]( \bar{c} \gamma^\mu b) \nnb \\ 
 &-& C_{A} [\bar{\ell}\gamma_\mu (1-\gamma_5) \nu_{\ell}]( \bar{c} \gamma^\mu \gamma_{5} b)\Big], 
\end{eqnarray}
where  $C_{V} $ and $C_{A}$ represent the Wilson coefficients including the SM contributions as well as those of  the operators coming from vector and pseudo-vector type of LQ interactions, respectively. At the $ \mu= M_{U}$ scale, $C_{V} $ and $C_{A}$ are written as
 \begin{eqnarray} \label{CVA} 
C_{V} = C_{A} &=& 1+ {\sqrt{2} g_{b\tau}^ * ({\cal V}_{g})_{c\tau}
 \over 4G_{F} V_{cb} M_{U}^2 } . 
\end{eqnarray}
   In $ \tau $, channel we use two optimal solutions, called $R_A$ and $R_B$,  obtained by the fitting of the parameters on the data in  $B\rightarrow D^{(*)}\ell \nu$ channel  \cite{Freytsis:2015qca,Zhang:2019jax,Mu:2019bin}. Ref. \cite{Freytsis:2015qca},  using  a general operator analysis,  identifies which four-fermion operators simultaneously fit  to $R(D)$ and $R(D^*)$ results. According to this paper, the values below provide us with the best fit values for the coefficients with acceptable $q^2$ spectra and $\chi_{min}^{2}<5$.  Obtained from these analyses the two values for  $g _{b\tau}^{*}({\cal V}_g)_{c\tau}  $ are given as \cite{Freytsis:2015qca}:
\begin{eqnarray} \label{CVA} 
g _{b\tau}^{*}({\cal V}_g)_{c\tau}= \left(\frac{M_U}{TeV}\right)^2 \left\{ \begin{array}{l} 
  ~~0.18\pm 0.04\qquad R_{A} \\[1.2ex]
  -2.88\pm 0.04 \qquad R_{B} \, 
                 \end{array} \right.,
\end{eqnarray}
where $ M_{U} $ is chosen  as $M_{U}=1 TeV$  at the scale $ \mu=M_{U}$ by taking into account the constraints on the vector LQ mass provided by CMS collaboration \cite{Chatrchyan:2012sv,Sirunyan}.   Although the fit results of $R_{A}$ and $R_{B}$ are quite different, the $C_{V}$ and $C_{A}$ coefficients have almost the same absolute values as $R_{A}$ and $R_{B}$ are entered with different signs. So, it is quite difficult to distinguish between  two results as they  lead to the same values for some physical observables. In the literature,  these best fit values are used in the analysis of many physical quantities associated with different semileptonic channels. In \cite{Li:2016pdv}, using the above best-fit solutions, the effects of  vector  LQs on some physical quantities defining  the semileptonic $\Lambda_{b}\rightarrow \Lambda_{c} \tau  \overline{\nu}_\tau$ channel are analyzed. The recent work \cite{Mu:2019bin}, investigates possible NP effects on the observables of $\Lambda_{b}\rightarrow \Lambda_{c} \tau  \overline{\nu}_\tau$ channel using the same fit values.  For more details on these parameters and their  effects on the physical quantities, see, for instance,  \cite{Freytsis:2015qca,Zhang:2019jax,Li:2016pdv,Mu:2019bin,Fajfer:2015ycq,Angelescu:2018tyl} and references therein.

 In Ref. \cite{Fajfer:2015ycq},  by attributing the difference between the experimental and indirect determinations of $ V_{cb} $ to the leptoquark contribution, the following constraint in $ \mu $ channel is obtained:
\begin{eqnarray} \label{mumu} 
 \mid V_{cb}\mid Re\Bigg(\frac{g _{b\mu}^{*}({\cal V}_g)_{c\mu}}{V_{cb}}\Bigg)\in [-0.1, -0.01]\times 10^{-3} (\frac{M_{U}}{TeV})^2,\nonumber\\
\end{eqnarray}
which will be used in our analyses. 

\section{The Transation Amplitude and  Form Factors}
The amplitude of the decay $  \Lambda_{b}  \rightarrow {\Lambda}_{c}  \ell \overline{\nu}_{\ell} $ is obtained  by sandwiching the effective Hamiltonian between the initial and final  baryonic states:
\begin{eqnarray}\label{amplitude}
{\cal M}^{ { \Lambda}_{b}  \rightarrow {\Lambda}_{c}  \ell \overline{\nu}_{\ell}} = \langle { \Lambda}_{c}, \lambda_2  \mid{\cal H}^{eff}_{SM+LQ}\mid 
{\Lambda}_{b}, \lambda_1 \rangle~,
\end{eqnarray}
where $\lambda_1$ and $\lambda_2$ are the helicities of the parent and daughter baryons, respectively.
The hadronic matrix elements of the axial and vector currents, inside the Hamiltonian,  are  parameterized by six hadronic form factors ($f_{1,2,3}$ and $g_{1,2,3}$) \cite{Shivashankara:2015cta,Gutsche:2015mxa}:
\begin{eqnarray}\label{SMtransmatrix} 
{\cal M}^{V}_{\mu}&=& \langle  \Lambda_c, \lambda_2 \mid  V^{\mu} \mid \Lambda_b, \lambda_1 \rangle 
= \bar {u}_{\Lambda_c} (p_2, \lambda_2)\Bigg[\gamma_{\mu}f_{1}(q^{2}) \nnb \\ 
&+&{i}\sigma_{\mu\nu}q^{\nu}f_{2}(q^{2})+ q^{\mu}f_{3}(q^{2})\Bigg] u_{\Lambda_b}(p_1, \lambda_1)  ,\nnb \\
\end{eqnarray}
and 
\begin{eqnarray}\label{SMtransmatrix} 
{\cal M}^{A}_{\mu}&=& \langle
\Lambda_c, \lambda_2 \mid  A^{\mu} \mid \Lambda_b, \lambda_1 \rangle 
=\bar {u}_{\Lambda_c} (p_2, \lambda_2) \Bigg[\gamma_{\mu}g_{1}(q^{2})\nnb \\
&+&{i}\sigma_{\mu\nu}q^{\nu}g_{2}(q^{2})+ q^{\mu}g_{3}(q^{2})\Bigg] \gamma_5  u_{\Lambda_b}(p_1, \lambda_1),\nnb \\
\end{eqnarray}
 where $\sigma_{\mu \nu}=\dfrac{i}{2}[\gamma_\mu,\gamma_\nu]$ and $q^{\mu}=(p_1 -p_2)^{\mu}$ is the four momentum transfer. Here, $V^{\mu}=\bar {c}\gamma_{\mu}b$ and $A^{\mu}=\bar {c}\gamma_{\mu}\gamma_5 b$ represent the vector and axial vector parts of the transition current, respectively and  $\bar {u}_{\Lambda_c} (p_2, \lambda_2)$ and $u_{\Lambda_b}(p_1, \lambda_1)$ are the corresponding  Dirac spinors for the final and initial baryonic states. The transition matrix elements can be parameterized in terms of the four-vector velocities $\upsilon_{\mu}$ and $\upsilon_{\mu}^{'}$, as well:
\begin{eqnarray}\label{SMtransmatrix} 
{\cal M}^{V}_{\mu}&=& \langle  \Lambda_c, \lambda_2 \mid  V^{\mu} \mid \Lambda_b, \lambda_1 \rangle 
= \bar {u}_{\Lambda_c} (p_2, \lambda_2)\Bigg[\gamma_{\mu}F_{1}(q^{2}) \nnb \\ 
&+&F_{2}(q^{2}) \upsilon_{\mu}+ F_{3}(q^{2}) \upsilon_{\mu}^{'}\Bigg] u_{\Lambda_b}(p_1, \lambda_1)  ,\nnb \\
\end{eqnarray}
and 
\begin{eqnarray}\label{SMtransmatrix} 
{\cal M}^{A}_{\mu}&=& \langle
\Lambda_c, \lambda_2 \mid  A^{\mu} \mid \Lambda_b, \lambda_1 \rangle 
=\bar {u}_{\Lambda_c} (p_2, \lambda_2) \Bigg[\gamma_{\mu}G_{1}(q^{2})\nnb \\
&+&G_{2}(q^{2}) \upsilon_{\mu}+ G_{3}(q^{2}) \upsilon_{\mu}^{'}\Bigg] \gamma_5  u_{\Lambda_b}(p_1, \lambda_1).\nnb \\
\end{eqnarray}
As we previously mentioned,  the form factors 
  $F_{1,2,3}$ and $G_{1,2,3}$ have been calculated   in full QCD and are available \cite{Azizi:2018axf}. The following relations describe the two sets of form factors in terms of each other (see also \cite{Azizi:2018axf,Shivashankara:2015cta,Gutsche:2015mxa,Zhang:2019jax}):
\begin{eqnarray} \label{f1} 
f_{1}(q)^2 &=& F_{1}(q)^2 + (m_{\Lambda_b} + m_{\Lambda_c}) \Bigg[ {F_2(q)^2  \over 2m_{\Lambda_b} } + {F_3(q)^2  \over 2m_{\Lambda_c} }   \Bigg], \nnb \\ 
f_{2}(q)^2 &=& {F_2(q)^2  \over 2m_{\Lambda_b} } + {F_3(q)^2  \over 2m_{\Lambda_c} }, \nnb \\ 
f_{3}(q)^2 &=& {F_2(q)^2  \over 2m_{\Lambda_b} } - {F_3(q)^2  \over 2m_{\Lambda_c}}, \nnb \\ 
g_{1}(q)^2 &=& G_{1}(q)^2 + (m_{\Lambda_c} - m_{\Lambda_b}) \Bigg[ {G_2(q)^2  \over 2m_{\Lambda_b} } + {G_3(q)^2  \over 2m_{\Lambda_c}}   \Bigg], \nnb \\ 
g_{2}(q)^2 &=& {G_2(q)^2  \over 2m_{\Lambda_b} } + {G_3(q)^2  \over 2m_{\Lambda_c} },  \nnb \\
g_{3}(q)^2 &=& {G_2(q)^2  \over 2m_{\Lambda_b} } - {G_3(q)^2  \over 2m_{\Lambda_c} }.
\end{eqnarray}
We would like to introduce the helicity amplitudes in terms of the various form factors and the NP couplings:
\begin{eqnarray} \label{HAHV} 
H_{\lambda_2 ,\lambda_W}^{V(A)} &=&\epsilon^{\dagger\mu} (\lambda_W) \langle \Lambda_c, \lambda_2 \mid V(A)^{\mu}\mid  \Lambda_b, \lambda_1 \rangle, \nnb \\ 
\mbox{and}\nnb \\
H_{\lambda_2 ,\lambda_W}&=&H_{\lambda_2 ,\lambda_W}^{V} - H_{\lambda_2 ,\lambda_W}^{A}.
\end{eqnarray}
where $\lambda_W$ indicates the helicity of $W^{-}_{off-shell}$. The expressions of the helicity amplitudes are defined as follows  \cite{Azizi:2018axf,Shivashankara:2015cta,Dutta:2015ueb,Zhang:2019jax}:
\begin{widetext}
\begin{eqnarray} \label{HAHV} 
H_{1/2,0}^V &=& {\sqrt{(m_{\Lambda_b} - m_{\Lambda_c})^2 -q^2} \over \sqrt{q^2} } [(m_{\Lambda_b} + m_{\Lambda_c}) f_1 (q^2) - q^2 f_2 (q^2)],\nnb \\ 
H_{1/2,0}^A &=&  {\sqrt{(m_{\Lambda_b} + m_{\Lambda_c})^2 -q^2} \over \sqrt{q^2} } [(m_{\Lambda_b} - m_{\Lambda_c}) g_1 (q^2) + q^2 g_2 (q^2)] ,\nnb \\ 
H_{1/2,1}^V &=& \sqrt{2[(m_{\Lambda_b} - m_{\Lambda_c})^2 -q^2]}  [- f_1 (q^2) + (m_{\Lambda_b} + m_{\Lambda_c}) f_2 (q^2)] ,\nnb \\ 
H_{1/2,1}^A &=&  \sqrt{2[(m_{\Lambda_b} + m_{\Lambda_c})^2 -q^2]}  [- g_1 (q^2) + (m_{\Lambda_b} - m_{\Lambda_c}) g_2 (q^2)], \nnb \\ 
H_{1/2,t}^V &=& {\sqrt{(m_{\Lambda_b} + m_{\Lambda_c})^2 -q^2} \over \sqrt{q^2} } [(m_{\Lambda_b} - m_{\Lambda_c}) f_1 (q^2) + q^2 f_3 (q^2)] ,\nnb \\ 
H_{1/2,t}^A &=& {\sqrt{(m_{\Lambda_b} - m_{\Lambda_c})^2 -q^2} \over \sqrt{q^2} } [(m_{\Lambda_b} + m_{\Lambda_c}) g_1 (q^2) - q^2 g_3 (q^2)], \nnb \\ 
\end{eqnarray}
\end{widetext}
where  $H_{\lambda_2 ,\lambda_W}^V = H_{- \lambda_2 , - \lambda_W}^V $ and $H_{\lambda_2 ,\lambda_W}^A = - H_{- \lambda_2 , - \lambda_W}^A $. We will use these helicity amplitudes to calculate the desired physical quantities in terms of hadronic form factors.

\section{Physical Observables}
Using the helicity amplitudes in terms of the hadronic transition form factors discussed in the previous section,  we would like to  introduce some physical observables defining the transition under consideration such as the differential decay width and  branching ratio,  the lepton forward-backward asymmetry  and 
$  R (\Lambda_c) $. Using the form factors  from full QCD, we discuss the behavior of these quantities with respect to $ q^2 $ and compare the SM predictions with those of the SM+LQ to search for possible shifts.
\subsection{The Differential Decay Width}
Making use of the amplitude and the  standard prescriptions, the differential angular distributions for $ \Lambda_b \rightarrow \Lambda_c \ell ~ \overline{\nu}_\ell$ decay channel can be written as \cite{Azizi:2018axf,Shivashankara:2015cta,Gutsche:2015mxa,Zhang:2019jax,Hu:2018veh}
\begin{eqnarray} \label{gamma} 
{d\Gamma (\Lambda_b \rar \Lambda_c \ell \overline{\nu}_{\ell} ) \over dq^2dcos\Theta_l} &=& { G_{F}^2 \vert V_{cb} \vert ^2  q^2  \vert \overrightarrow{p}_{\Lambda_{c}} \vert \over 512 \pi^3 m_{\Lambda_{b}}^2 } 
\Bigg(1- {m_l^2 \over q^2} \Bigg)^2 \nnb \\  
&&\Bigg[  A_1 + {m_l^2 \over q^2} A_2  \Bigg],
\end{eqnarray}
where
\begin{eqnarray} \label{gamma1} 
A_1 &=& C_{V}^2 [2 sin^2\Theta_l (H_{1/2,0}^2 + H_{-1/2,0}^2) + (1 - cos\Theta_l)^2\nnb \\ 
&& H_{1/2,1}^2 + (1 + cos\Theta_l)^2 H_{-1/2,-1}^2] ,\nnb \\ 
A_2 &=& C_{V}^2 [2 cos^2\Theta_l (H_{1/2,0}^2 + H_{-1/2,0}^2) + sin^2\Theta_l  \nnb \\ 
&&(H_{1/2,1}^2 + H_{-1/2,-1}^2) + 2 (H_{1/2,t}^2 + H_{-1/2,t}^2) \nnb \\ 
&-& 4 cos\Theta_l (H_{1/2,t} H_{1/2,0} + H_{-1/2,t} H_{-1/2,0}) ] ,\nnb \\ 
\vert \overrightarrow{p}_{\Lambda_{c}} \vert &=& {\sqrt{\Delta } \over 2m_{ \Lambda_b} },\nnb \\ 
\Delta&=& (m_{ \Lambda_b}^2)^2 + (m_{ \Lambda_c}^2)^2 + (q^2)^2 - 2(m_{ \Lambda_b}^2 m_{ \Lambda_c}^2 \nnb \\ 
&+& m_{ \Lambda_c}^2 q^2 + m_{ \Lambda_b}^2 q^2). ~~\nnb \\ 
\nnb \\ 
\end{eqnarray}
Here  $\Theta_l$  indicates  the  angle between  momenta   of the lepton  and the baryon ${\Lambda}_c$ in the $q^2$ rest frame.

\subsection{The Differential Branching Ratio}
In this subsection, we perform the numerical analysis of the differential branching ratio and discuss its dependence on  $q^2$ at the $ \mu $ and $ \tau $ channels. To this end, we need   the values of some input parameters presented  in Table I \cite{PDG}.
\begin{table}
\begin{tabular}{|c|c|}
\hline\hline
Some Input Parameters & Values \\ \hline\hline
$ m_{\Lambda_b} $    &   $ 5.6196               $   $GeV$ \\
$ m_{\Lambda_c} $      &   $ 2.2864              $   $GeV$ \\
$ \tau_{\Lambda_b} $ &   $ 1.47\times 10^{-12} $   $s$      \\
$ G_{F} $            &   $ 1.166\times 10^{-5}  $   $GeV^{-2}$ \\
$ | V_{cb}|$ &   $ 0.0422               $                \\
$ m_{\mu} $              &   $ 0.1056      $   $GeV$ \\
$ m_{\tau} $              &   $ 1.7768          $   $GeV$ \\
 \hline \hline
\end{tabular}%
\caption{The values of some input parameters  used in our calculations \cite{PDG}. Note that in this table we show only the central values of the input parameters, while we take into account their uncertainties in the numerical calculations, as well.}
\label{tab:Param}
\end{table}
Also, we need the fit functions of the form factors calculated via light cone QCD sum rules in full theory as the main inputs in SM and BSM. As we previousely mentioned,  these fits are available in Ref. \cite{Azizi:2018axf}.  They are given in terms of $ q^2 $ as
\begin{eqnarray} \label{gamma1} 
{\cal F}(q^2)=\dfrac{{\cal F}(0)}{(1-\xi_1 \dfrac{q^2}{m_{\Lambda_b}^{2}}+\xi_2 \dfrac{q^4}{m_{\Lambda_b}^{4}}+\xi_3 \dfrac{q^6}{m_{\Lambda_b}^{6}}+\xi_4 \dfrac{q^8}{m_{\Lambda_b}^{8}})}
\end{eqnarray} 
where the $\xi_1 $, $\xi_2 $, $\xi_3 $ and $\xi_4$ are  fit parameters; and ${\cal F}(0)$ denotes the value of related form factor at $q^2=0$. 
The numerical values of these parameters are presented in Table II.
\begin{table}
\begin{tabular}{cclcccc}
\hline\hline
$Form factors$        & ${\cal F}(q^2=0)$               & $\xi_1$ &  $\xi_2$ & $\xi_3$ & $\xi_4$ \\ \hline \hline
$ F_{1} (q^2) $    &   $ 1.220\pm0.293   $  & ~$1.03$  & ~$-4.60 $ &  ~$28$ & ~$-53$ \\
$ F_{2} (q^2) $    &  $ -0.256\pm0.061   $   & ~$2.17$  & ~$-8.63 $ &  ~$51.40$ & ~$-85.2$ \\
$ F_{3} (q^2) $    &   $-0.421\pm0.101   $   & ~$2.18$  & ~$-1.02 $ &  ~$18.12$ & ~$-32$\\
$ G_{1} (q^2) $    &   $ 0.751\pm0.180   $   & ~$1.41$  & ~$-3.30 $ &  ~$21.90$ & ~$-40.10$ \\
$ G_{2} (q^2) $    &   $-0.156\pm0.037   $   & ~$1.46$  & ~$-6.50 $ & ~ $41.20$ & ~$-74.82$ \\
$ G_{3} (q^2)  $   &   $ 0.320\pm0.077   $    & ~$2.36$  & ~$-2.90 $ & ~ $28.20$ & ~$-45.20$ \\
 \hline \hline
\end{tabular}%
\caption{Parameters of the fit functions for different form factors  for $\Lambda_b \rar \Lambda_c $ decay \cite{Azizi:2018axf}.}
\label{tab:Param}
\end{table}

The differential branching ratio  as a function of  $q^2$ is obtained as
\begin{eqnarray}
DBR(q^2)=\Bigg(  \int_{-1}^{1}   \dfrac{d\Gamma(\Lambda_{b}\rightarrow \Lambda_{c} \ell \overline{\nu}_\ell)}{dq^2 dcos\Theta_l} d cos\Theta_l \Bigg) / \Gamma_{tot},~
\end{eqnarray}
where $ \Gamma_{tot}=\frac{\hbar}{ \tau_{\Lambda_b} } $.
In order to see how the predictions of the vector LQ model deviate from those of the SM, we plot the differential branching ratio of  $\Lambda_{b}\rightarrow \Lambda_{c} \ell \overline{\nu}_\ell$ transition at $\mu$ and $\tau$ channels in the SM and vector LQ models in figures 1 and 2. Figure 1 belongs to the $DBR(q^2)-q^2  $ at $ \mu $ channel including all errors coming from the LQ model parameters,  form factors as well as other input parameters. Note that the main errors belong to the uncertainties of the form factors and the errors coming from the LQ model parameters are very small at $\mu$ channel. This figure also contains the data provided by the LHCb Collaboration  \cite{Aaij:2017svr}.  As it  is seen,   the LQ and SM have exactly the same predictions on differential branching ratio at $\mu$ channel and they include the data.   The $q^2$-behavior of the $ DBR $ in both models is consistent with the data, such that $ DBR $ increases with the increasing of the $q^2$ and after reaching to a maximum it starts to decrease.

\begin{figure}[h!]
\centering
\begin{tabular}{ccc}
\epsfig{file=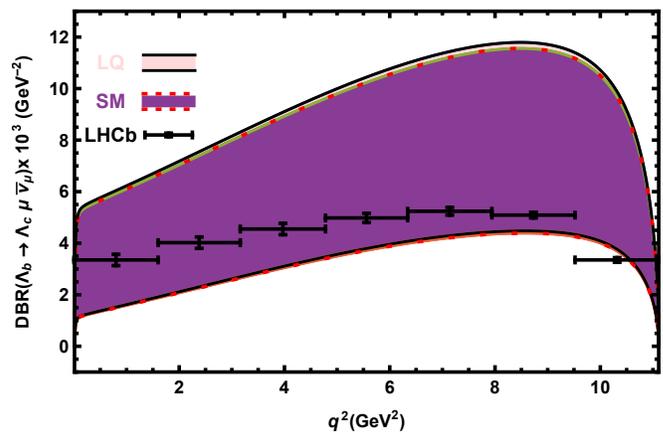,width=1.0\linewidth,clip=} &
\end{tabular}
\caption{The dependence of the $ DBR $ on  $q^2$  for the $\Lambda_{b}\rightarrow \Lambda_{c} \mu  \overline{\nu}_\mu$  transition  in the SM and vector LQ models  with all errors.  The experimental data have been taken from the LHCb Collaboration,  Ref. \cite{Aaij:2017svr}.}
\end{figure}

\begin{widetext}

\begin{figure}[h!]
\begin{center}
\includegraphics[totalheight=6cm,width=8cm]{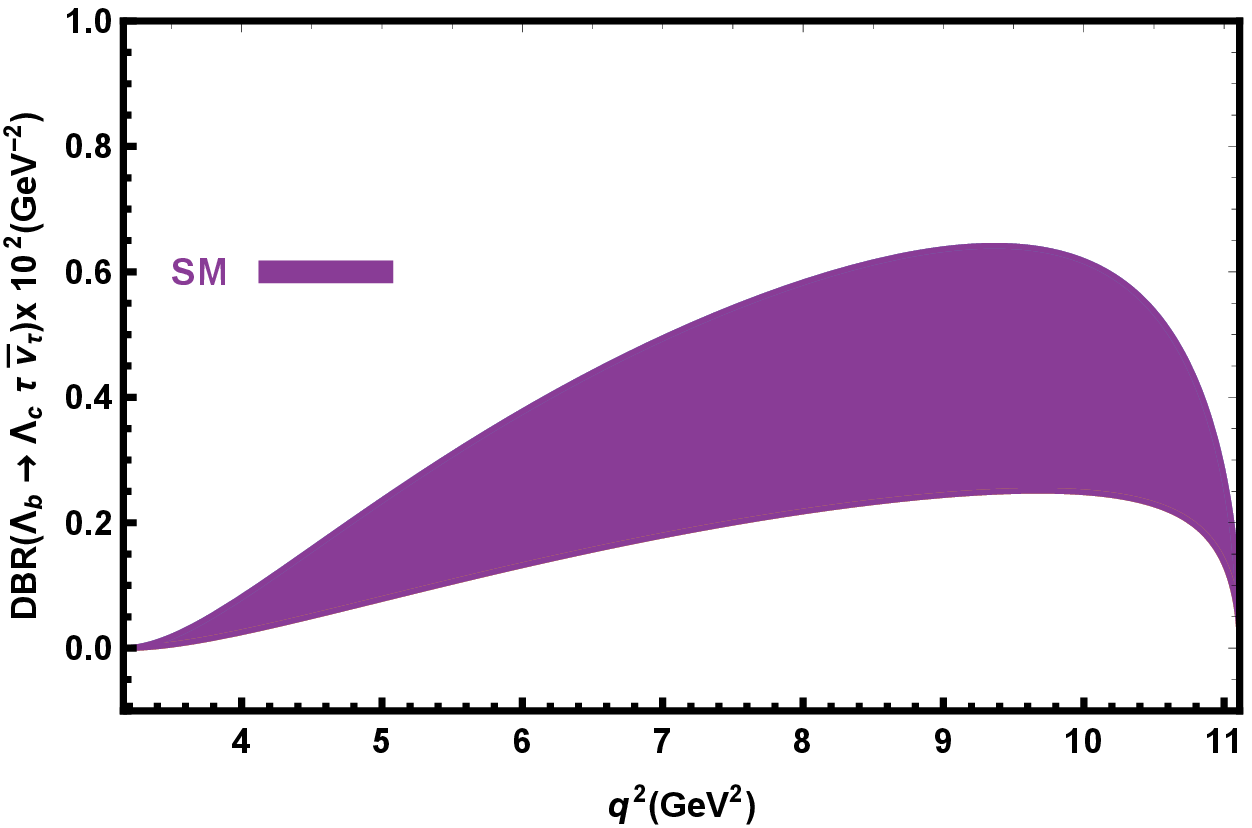}
\includegraphics[totalheight=6cm,width=8cm]{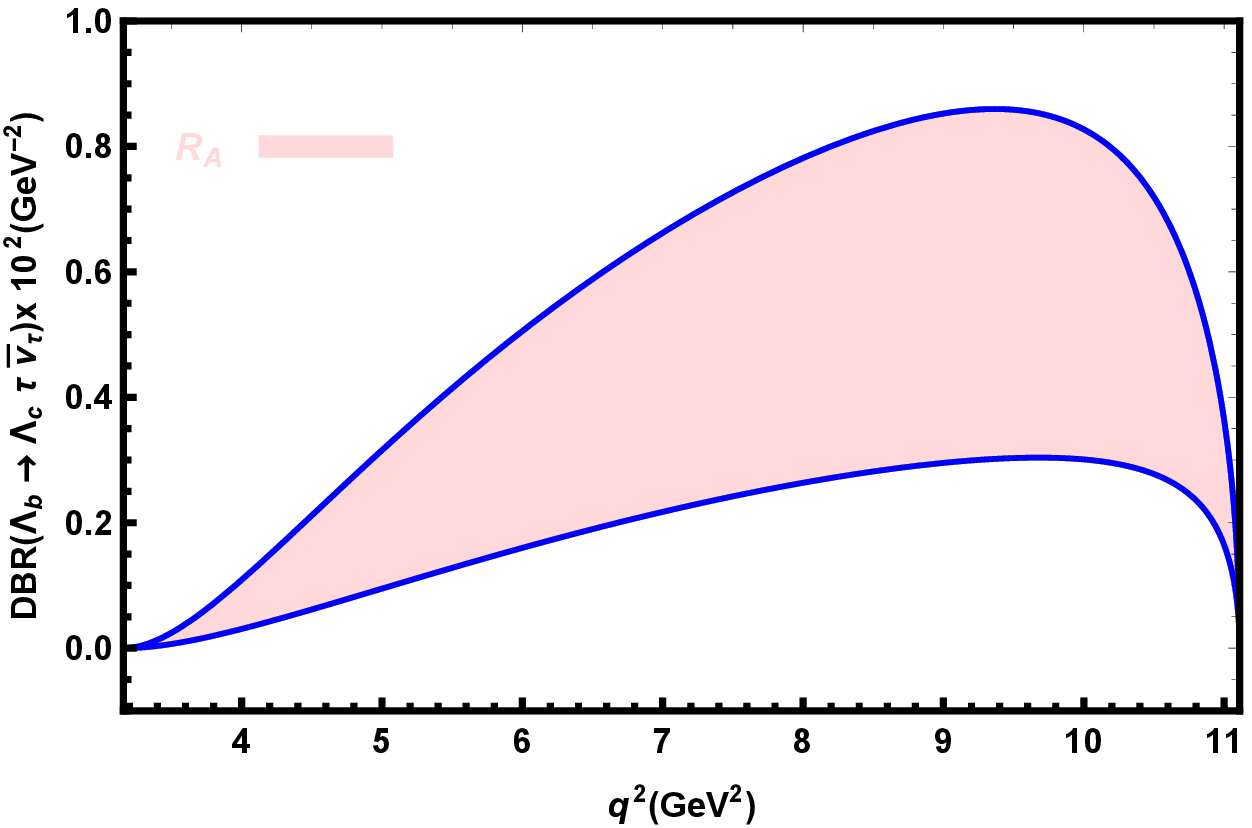}
\end{center}
\end{figure}
\begin{figure}[h!]
\begin{center}
\includegraphics[totalheight=6cm,width=8cm]{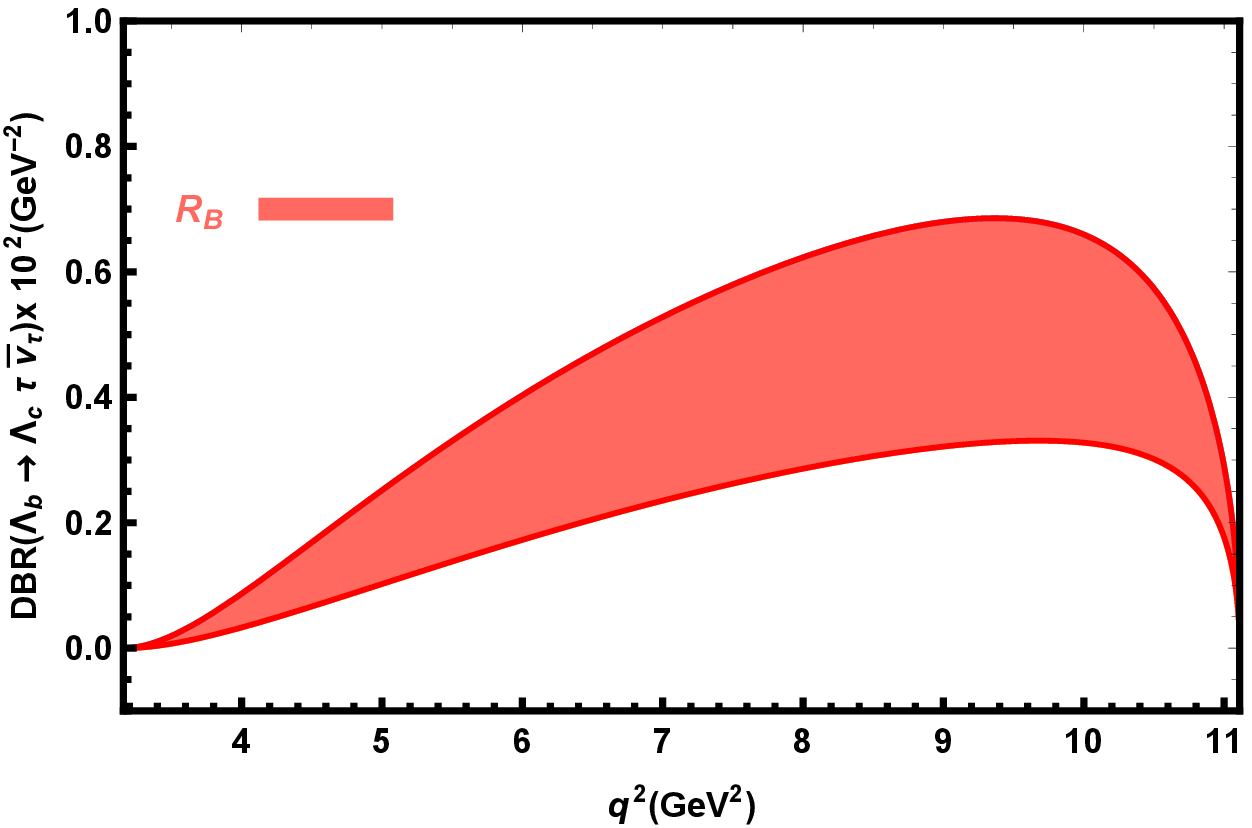}
\includegraphics[totalheight=6cm,width=8cm]{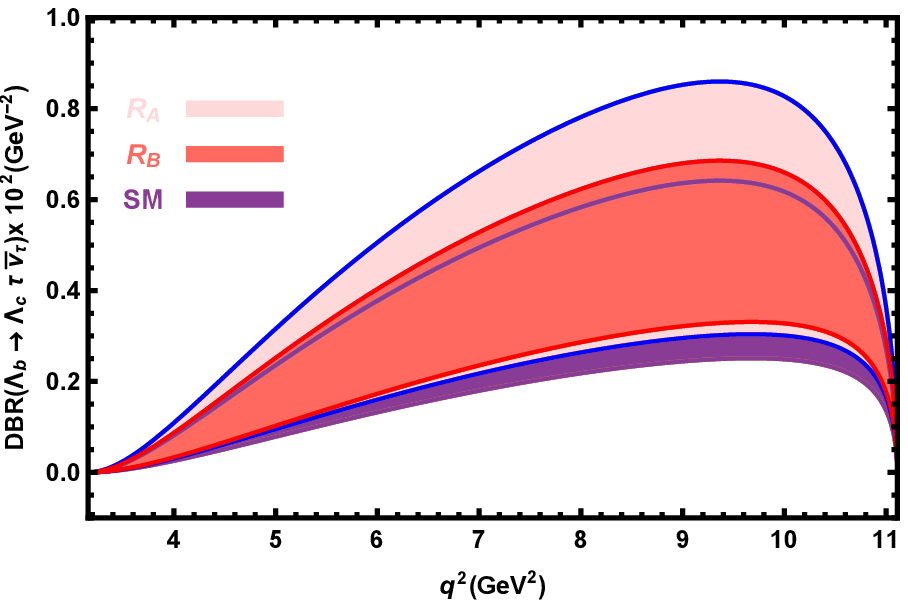}
\end{center}
\caption{The dependence of the $ DBR $ on  $q^2$  for the  $\Lambda_{b}\rightarrow \Lambda_{c} \tau  \overline{\nu}_\tau$ transitions  in the SM and vector LQ models  (separately and all together) with all errors.}
\end{figure}

\end{widetext}

As far as the $DBR(q^2)-q^2  $ at $ \tau $ channel is concerned, figure 2 shows that there are considerable deviations of  $R_A$  type LQ model predictions from the SM band. The band of  $R_B$  type LQ model  shows a shift from the  SM band,  as well but the violation in this case  is relatively small compared to the   $R_A$  type LQ model. 

In table III, we  present  the branching ratios in $\mu$ and   $\tau$ channels obtained in SM and  LQ scenarios. We also present the experimental data from  PDG available for  $\Lambda_{b}\rightarrow \Lambda_{c} \mu  \overline{\nu}_\mu$ transition as well as the predictions of Ref. \cite{Li:2016pdv} in $ \tau $ channel.  We see that the SM and LQ predictions  in $\mu$  channel are nicely consistent with the experimental value. Note that as we also mentioned in the case of differential branching ratio, the SM and LQ have the same  predictions for the branching ratio in  $\mu$ channel. However, in $\tau$  channel as also discussed in the case of differential branching ratio, the prediction of both $R_A$ and  $R_B$  type LQ models  differ from the SM result, considerably although the violation for the  $R_A$ type model is relatively large.  The presented values of the branching ratios in  Ref. \cite{Li:2016pdv} and $ \tau $ channel  have been obtained using the form factors calculated via lattice QCD with 2 + 1 dynamical flavors in HQET limit. When we compare our results with those of Ref. \cite{Li:2016pdv}, we see a  consistency among our results on the branching ratios of the $\Lambda_{b}\rightarrow \Lambda_{c} \tau^{-}  \overline{\nu}_\tau  $ transition in both SM and vector LQ scenarios with those of Ref. \cite{Li:2016pdv} witin the presented errors. 


\begin{widetext}

\begin{table}
\begin{tabular}{|c|c|c|c|}
\hline\hline
                           & {Present Work (\%)}& {Exp. \cite{PDG}(\%)}& {Ref. \cite{Li:2016pdv} (\%)} \\ \hline\hline
{$BR^{SM}(\Lambda_{b}\rightarrow \Lambda_{c} \mu^{-}  \overline{\nu}_\mu)$ }  & {  $5.89^{+2.22}_{-1.14}$  } & {$6.2^{+1.4}_{-1.3}$} & - \\
{$ BR^{SM}(\Lambda_{b}\rightarrow \Lambda_{c} \tau^{-}  \overline{\nu}_\tau) $}  &{  $ 1.86^{+0.70}_{-0.32}$} &  - & {  $ 1.77^{+0.09}_{-0.09}$} \\
{$ BR^{LQ}(\Lambda_{b}\rightarrow \Lambda_{c} \mu^{-}  \overline{\nu}_\mu) $}   & { $   5.89^{+2.22}_{-1.14}$ }   &  - & - \\
{$ BR^{LQ}_{R_A}(\Lambda_{b}\rightarrow \Lambda_{c} \tau^{-}  \overline{\nu}_\tau) $}  & {  $2.38^{+0.98}_{-0.44}$ } &  - & {  $ 2.27^{+0.17}_{-0.17}$}\\
{$ BR^{LQ}_{R_B}(\Lambda_{b}\rightarrow \Lambda_{c} \tau^{-}  \overline{\nu}_\tau) $  } &  { $2.10^{+0.69}_{-0.24}$} &  - &{  $ 2.24^{+0.17}_{-0.17}$}\\
 \hline \hline
\end{tabular}%
\caption{The values of branching ratios for $\Lambda_{b}\rightarrow \Lambda_{c} \mu^{-}  \overline{\nu}_\mu $ and $\Lambda_{b}\rightarrow \Lambda_{c} \tau^{-}  \overline{\nu}_\tau  $ transitions.}
\label{tab:BranchingR}
\end{table}

\end{widetext}

\subsection{The Lepton Forward-Backward Asymmetry}
In this subsection, we would like to  deal with the lepton forward-backward asymmetry ($\textit{A}_{FB}$), which is  one of the important parameters sensitive to the new physics. It is defined as
\begin{widetext}
\begin{eqnarray}
A_{FB}(q^2)= \dfrac{\int_{0}^{1} \dfrac{d\Gamma}{d q^2 d cos\Theta_l } d cos\Theta_l - \int_{-1}^{0} \dfrac{d\Gamma}{d q^2 d cos\Theta_l } d cos\Theta_l }{\int_{0}^{1} \dfrac{d\Gamma}{d q^2 d cos\Theta_l } d cos\Theta_l +\int_{-1}^{0} \dfrac{d\Gamma}{d q^2 d cos\Theta_l } d cos\Theta_l }.
\end{eqnarray}
\end{widetext}
We plot the dependence of the lepton forward-backward asymmetry on $q^2$  at $\mu$ and $\tau$ channels both in the SM and vector LQ model in figures 3 and 4 considering all the encountered errors in the calculations. From these figures, we conclude that the  two models have predictions  in a roughly good consistency for all the possible cases at all lepton channels.  In the case of $\mu$, the $\textit{A}_{FB}$ changes its sign at very small values of $ q^2 $, while this point is shifted to the average values of $ q^2 $ at $\tau$ case. Any future data on the values and signs  of $\textit{A}_{FB}$ at different lepton channels and their comparison with the  predictions of the present study would help us get useful knowledge on the decay modes under study and the  internal structures of the participated baryons as well as restrict the parameters of the models BSM.

\begin{figure}[h!]
\centering
\begin{tabular}{ccc}
\epsfig{file=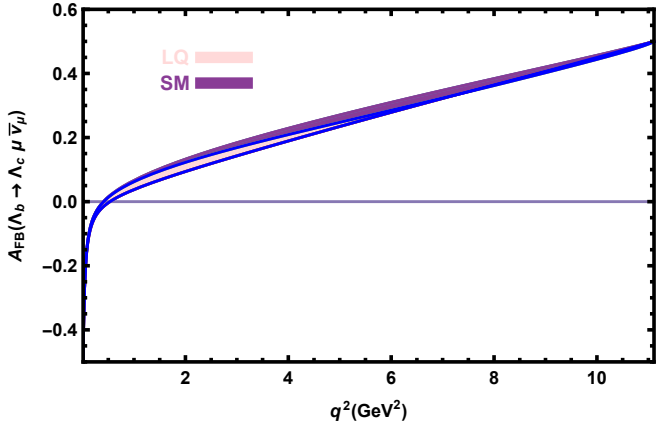,width=1.0\linewidth,clip=} 
\end{tabular}
\caption{The dependence of the $ A_{FB} $ on  $q^2$  for the $\Lambda_{b}\rightarrow \Lambda_{c} \mu  \overline{\nu}_\mu$  transition   in  SM and vector LQ models with all errors.}
\end{figure}
\begin{figure}[h!]
\centering
\begin{tabular}{ccc}
\epsfig{file=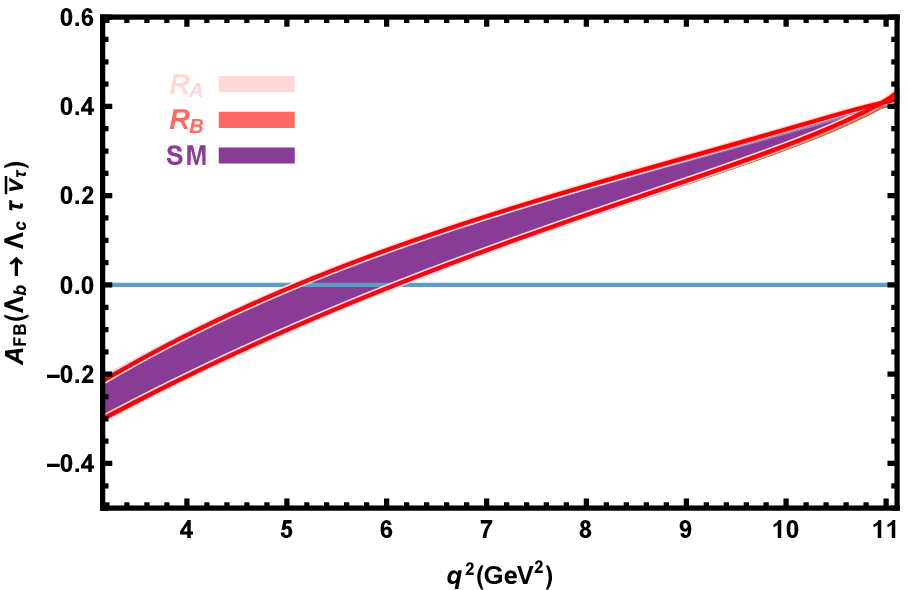,width=1.0\linewidth,clip=} 
\end{tabular}
\caption{The dependence of the $ A_{FB} $ on  $q^2$  for the $\Lambda_{b}\rightarrow \Lambda_{c} \tau  \overline{\nu}_\tau$  transition   in  SM and vector LQ models with all errors.}
\end{figure}
\begin{widetext}

\begin{figure}[h!]
\begin{center}
\includegraphics[totalheight=6cm,width=8cm]{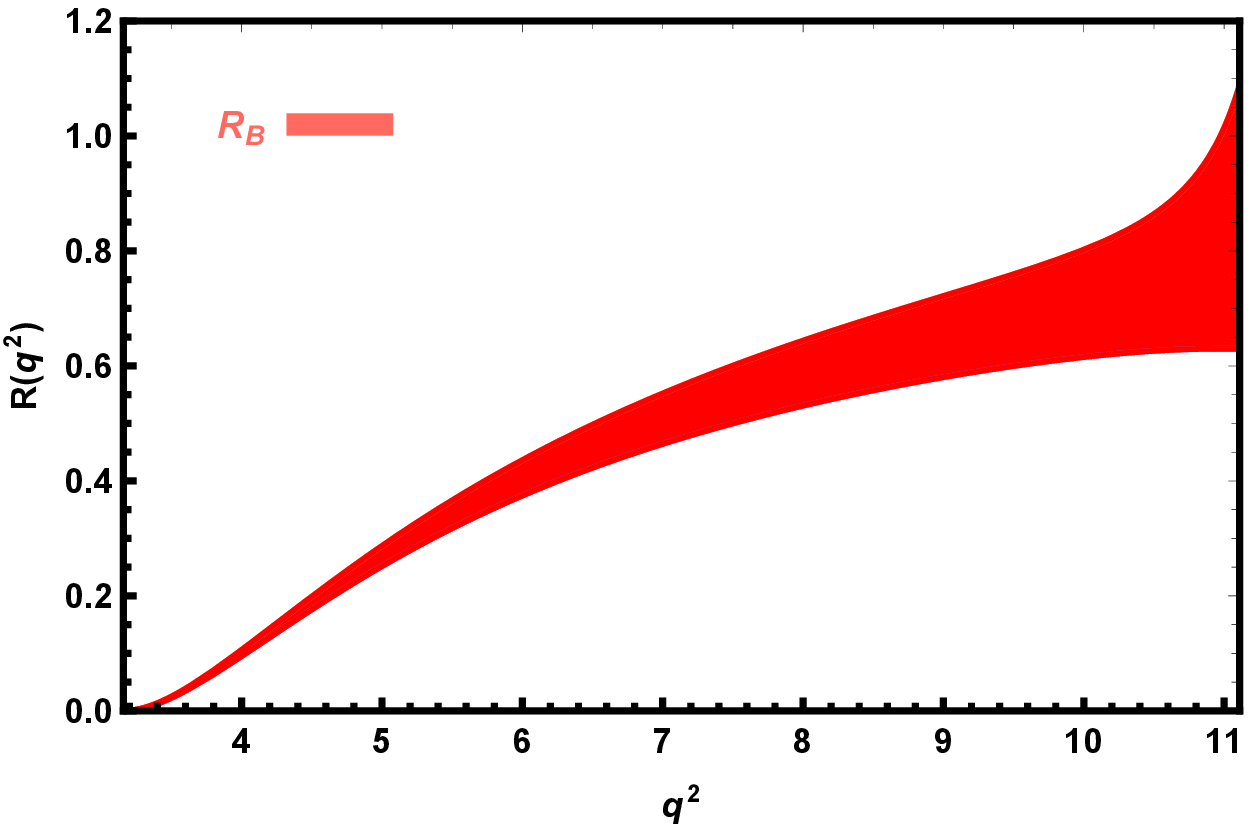}
\includegraphics[totalheight=6cm,width=8cm]{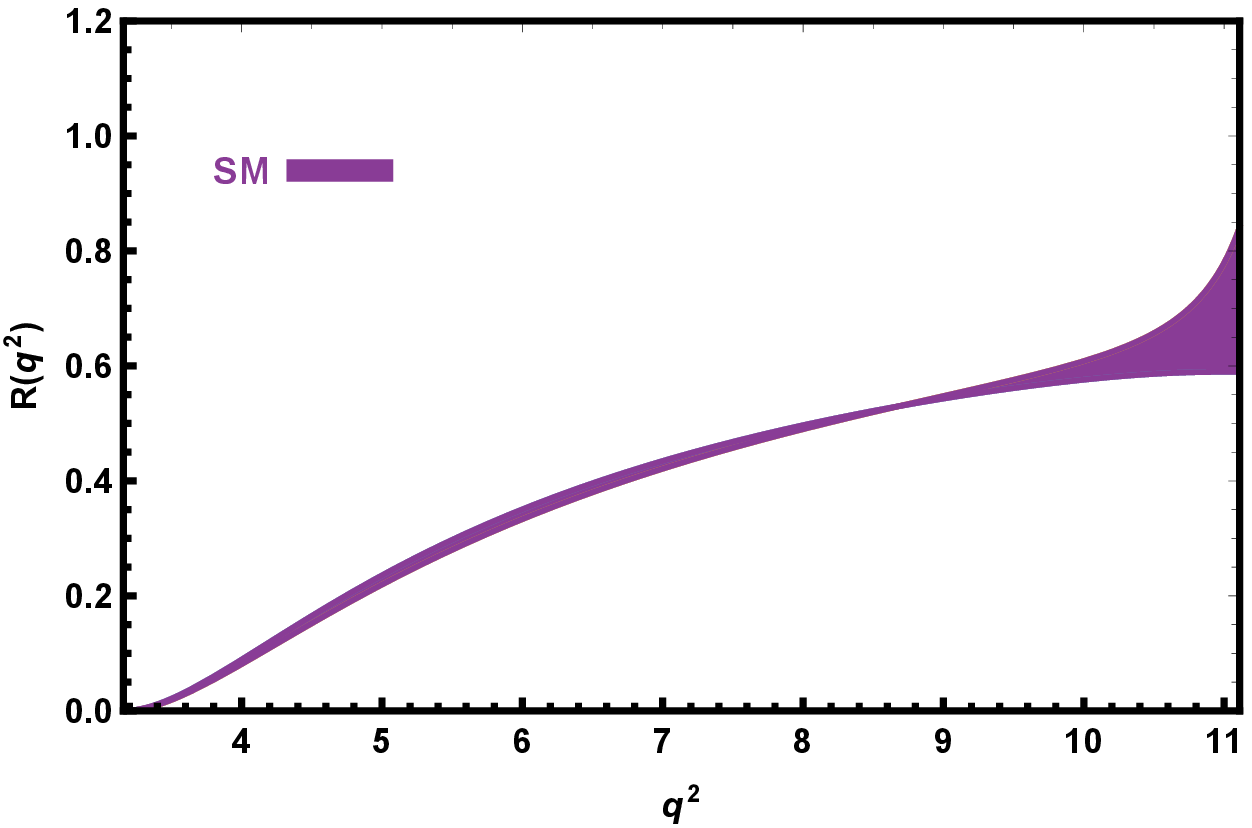}
\end{center}
\end{figure}
\begin{figure}[h!]
\begin{center}
\includegraphics[totalheight=6cm,width=8cm]{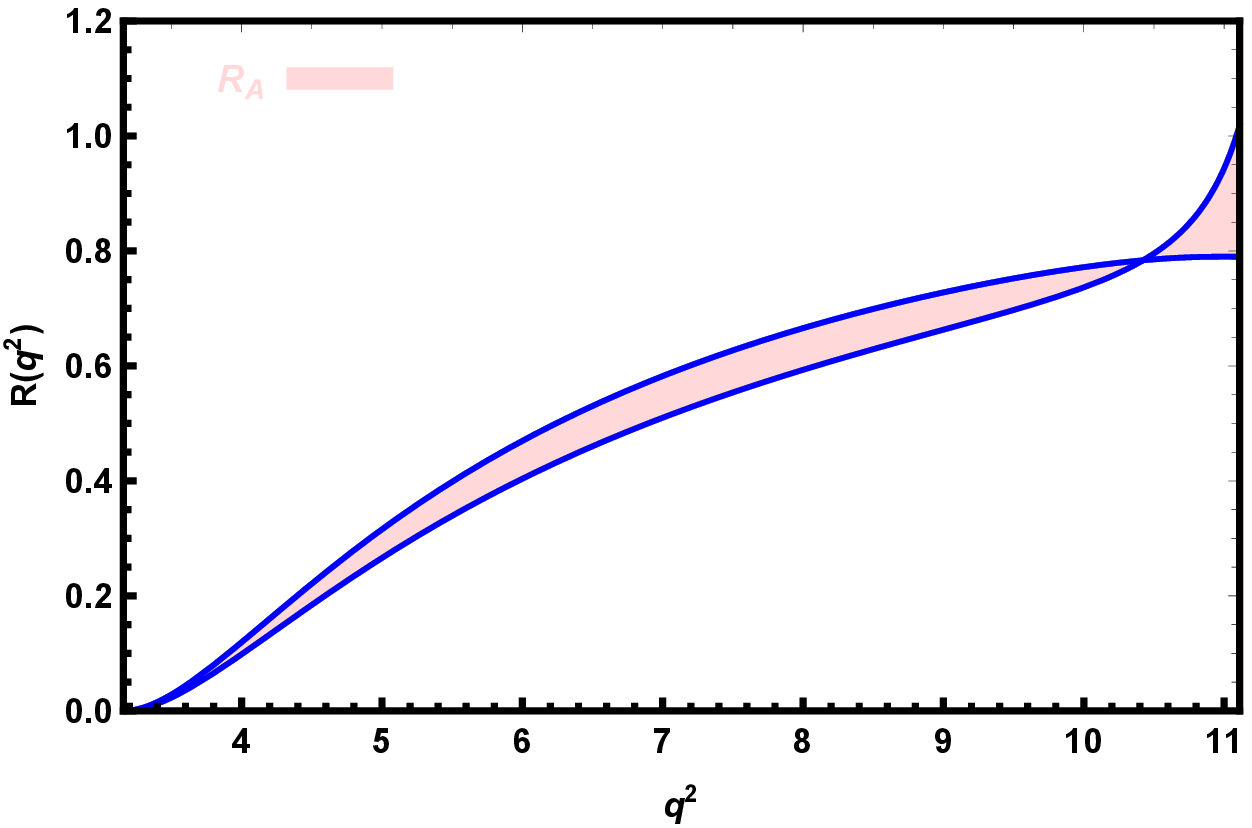}
\includegraphics[totalheight=6cm,width=8cm]{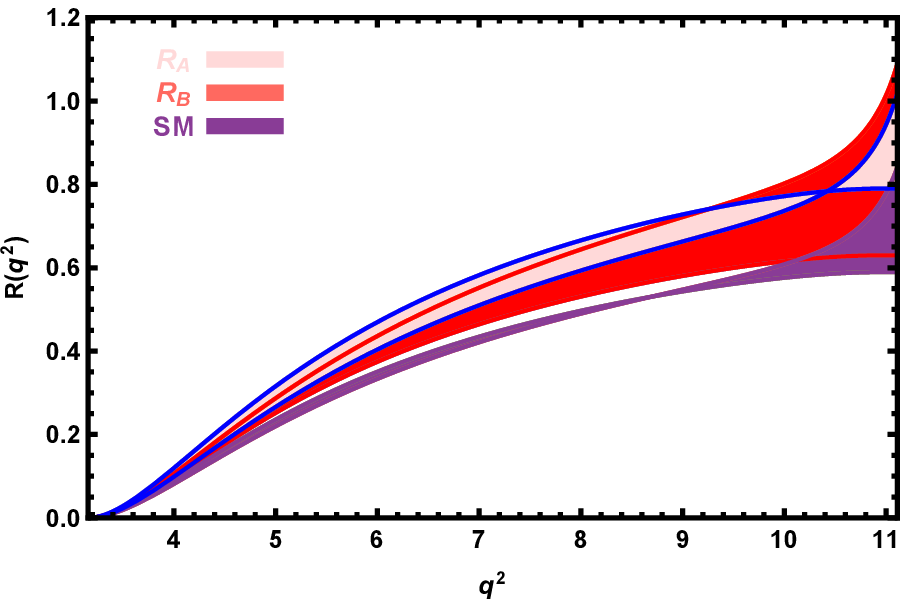}
\end{center}
\caption{The dependence of the $ R(q^2) $ on  $q^2$ in  SM and vector LQ models (separately and all together)  with all errors.}
\end{figure}

\end{widetext}
\subsection{The  Parameter $  R(q^2) $}
In this part, we  present the results for the ratio  of differential branching ratios in $\tau$  and $\mu$ channels, i. e., 
\begin{eqnarray} \label{R} 
R(q^2) =  \dfrac{DBR(q^2)(\Lambda_{b}\rightarrow \Lambda_{c} \tau  \overline{\nu}_\tau)}{DBR(q^2)(\Lambda_{b}\rightarrow \Lambda_{c} \mu\overline{\nu}_\mu)},
\end{eqnarray}
which is one of the most important probes to search for new physics effects. The experiments have shown serious  deviations from the SM predictions on this parameter in  some mesonic channels and we are witnessing serious violations of the lepton flavor universality in mesonic channels. The $  \Lambda_{b}  \rightarrow {\Lambda}_{c}  \ell \overline{\nu}_{\ell} $ is one of the important tree-level baryonic transitions, which is accessible in the experiments like LHCb. Testing the experimental data on $ R(\Lambda_c) $ and their comparison with the theoretical predictions are of great importance. 
We plot the dependence of the $R(q^2)$ on $q^2$ in the SM and vector LQ model in figure 5. From this figure, we see that the results obtained using   both the  $ R_A $  and $ R_B $ types  fit solutions in LQ model deviate from the SM predictions, considerably. Only at higher  values of $ q^2 $, the  $ R_B $ type  fit solution shows some intersections with the SM predictions.

It will be  instructive to give the values for $ R(\Lambda_c) $ both in SM and LQ scenarios, as well. By performing the integrals over $ q^2 $ in the allowed limits we find
\begin{eqnarray} 
R(\Lambda_c) &=& \frac{{ \cal B}(\Lambda_b\rightarrow \Lambda_c \tau \overline{\nu}_{\tau})}{{\cal B}(\Lambda_b\rightarrow \Lambda_c \mu\overline{\nu}_{\mu})} \nonumber\\ &=& \left\{ \begin{array}{l} 
(0.314-0.339) \quad SM\\
 (0.410-0.421)\quad  LQ~~ R_{A} \\
  (0.335-0.445)\quad  LQ ~R_{B} \, 
                 \end{array} \right. . \nonumber\\
\end{eqnarray}
From the obtained results we conclude that for both  the  $ R_A $ and  $ R_B $  types solutions,  LQ model predictions deviate from the SM prediction, considerably.  The band related to the  $ R_B $  type LQ model  shows only a very small overlap with the SM predictions.  We compare our results  on $R(\Lambda_c)$  with the predictions of  Ref. \cite{Li:2016pdv} in Table \ref{tab:R}. 
From this table, it is clear that our results  and the prediction of Ref. \cite{Li:2016pdv}  on the $R(\Lambda_c)$  are close to each other and the presented regions indicate some overlaps. 
\begin{table}
\begin{tabular}{|c|c|c|}
\hline\hline
                           & Present Work &  Ref. \cite{Li:2016pdv} \\ \hline\hline
$R(\Lambda_c )^{SM}$                     &    $0.314 - 0.339 $      &$ 0.320 - 0.340$ \\
$ R(\Lambda_c )^{LQ}$ (for $R_A$)    &   $ 0.410 - 0.421 $   &  $ 0.410 - 0.450$ \\
$ R(\Lambda_c )^{LQ}$ (for $R_B$)        &   $  0.335 - 0.445$    &    $  0.400 - 0.440$ \\
 \hline \hline
\end{tabular}%
\caption {Our results on $R(\Lambda_c )$  in comparison with  the predictions of Ref. \cite{Li:2016pdv}.}
\label{tab:R}
\end{table}
Future experimental data will tel us whether there are LFUV in  $  \Lambda_{b}  \rightarrow {\Lambda}_{c}  \ell \overline{\nu}_{\ell} $  channel or not. 

\subsection{Longitudinal Polarization of $\Lambda_{c}$ Baryon and  $ l $ Lepton }
In this subsection, we would like to present the $\Lambda_{c}$ baryon and lepton ($\mu$ and $\tau$) polarizations, which are important parameters to search for  new physics effects. 
These parameters  are defined as
\begin{eqnarray}
P_{\Lambda_c}(q^2)=\frac{{\rm d}\Gamma^{\lambda_2=1/2}/{\rm d}q^2-
    {\rm d}\Gamma^{\lambda_2=-1/2}/{\rm d}q^2}{{\rm d}\Gamma/{\rm d}q^2},\,
\end{eqnarray}
and
\begin{eqnarray}
 P_{\ell}(q^2)=\frac{{\rm d}\Gamma^{\lambda_{\ell}=1/2}/{\rm d}q^2-
    {\rm d}\Gamma^{\lambda_{\ell}=-1/2}/{\rm d}q^2}{{\rm d}\Gamma/{\rm d}q^2}\,.
\end{eqnarray}
where
\begin{eqnarray}
\frac{{\rm d}\Gamma^{\lambda_2=1/2}}{{\rm d}q^2}&=&\frac{m_\ell^2}{q^2}\Big[\frac{4}{3}C_V^2\big(H_{1/2,1}^2+H_{1/2,0}^2+
3H_{1/2,t}^2\big)\Big]\nnb \\ 
&&+\frac{8}{3}C_V^2\big(H_{1/2,0}^2+H_{1/2,1}^2\big),\nnb \\ 
\frac{{\rm d}\Gamma^{\lambda_2=-1/2}}{{\rm d}q^2}&=&\frac{m_\ell^2}{q^2}\Big[\frac{4}{3}C_V^2\big(H_{-1/2,-1}^2\!+\!H_{-1/2,0}^2\nnb \\ 
&&
+3H_{-1/2,t}^2\big)\Big]+\frac{8}{3}C_V^2\big(H_{-1/2,-1}^2+H_{-1/2,0}^2\big),\nnb \\ 
\frac{{\rm d}\Gamma^{\lambda_\ell=1/2}}{{\rm d}q^2}&=&\frac{m_\ell^2}{q^2}C_V^2\Big[\frac{4}{3}\big(H_{1/2,1}^2\!+\!H_{1/2,0}^2\!+\!
H_{-1/2,-1}^2\!\nnb \\ 
&&+\!H_{-1/2,0}^2\big)\!+\!4\big(H_{1/2,t}^2\!+\!H_{-1/2,t}^2\big)\Big],\nnb \\ 
\frac{{\rm d}\Gamma^{\lambda_\ell=-1/2}}{{\rm d}q^2}&=&\frac{8C_V^2}{3}\big(H_{1/2,1}^2\!+\!H_{1/2,0}^2\!+\!H_{-1/2,-1}^2\!\nnb \\ 
&&+\!H_{-1/2,0}^2\big)\,.
\end{eqnarray}

The dependence of $\Lambda_{c}$ baryon and  lepton polarizations on $q^2$ at $\mu$ and
$\tau$ channels in SM and vector LQ models with all errors  are presented in figures 6, 7, 8 and 9. From these figures, we observe  that the LQ and SM predictions show considerable differences on $ P_{\Lambda_{c}} -q^2$ in $ \mu $ channel. However, in $ \tau $ channel the SM  and both the LQ scenarios have  roughly the same predictions on $ P_{\Lambda_{c}}-q^2 $. For the case of $ P_{\mu} $, we see small shifts in some regions between the SM and LQ predictions. As far as the $ P_{\tau} $ is concerned, the $ R_A $ and  $ R_B $ type LQ scenarios have almost the same predictions, but their results show considerable deviations from the SM prediction.
\begin{figure}[h!]
\centering
\begin{tabular}{ccc}
\epsfig{file=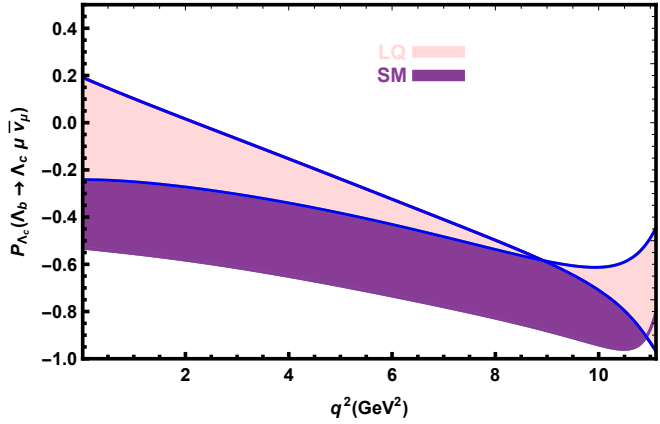,width=1.0\linewidth,clip=} 
\end{tabular}
\caption{The dependence of the $ P_{\Lambda_{c}} $ on  $q^2$  for the $\Lambda_{b}\rightarrow \Lambda_{c} \mu  \overline{\nu}_\mu$  transition   in  SM and vector LQ models with all errors.}
\end{figure}

\begin{widetext}

\begin{figure}[h!]
\begin{center}
\includegraphics[totalheight=6cm,width=8cm]{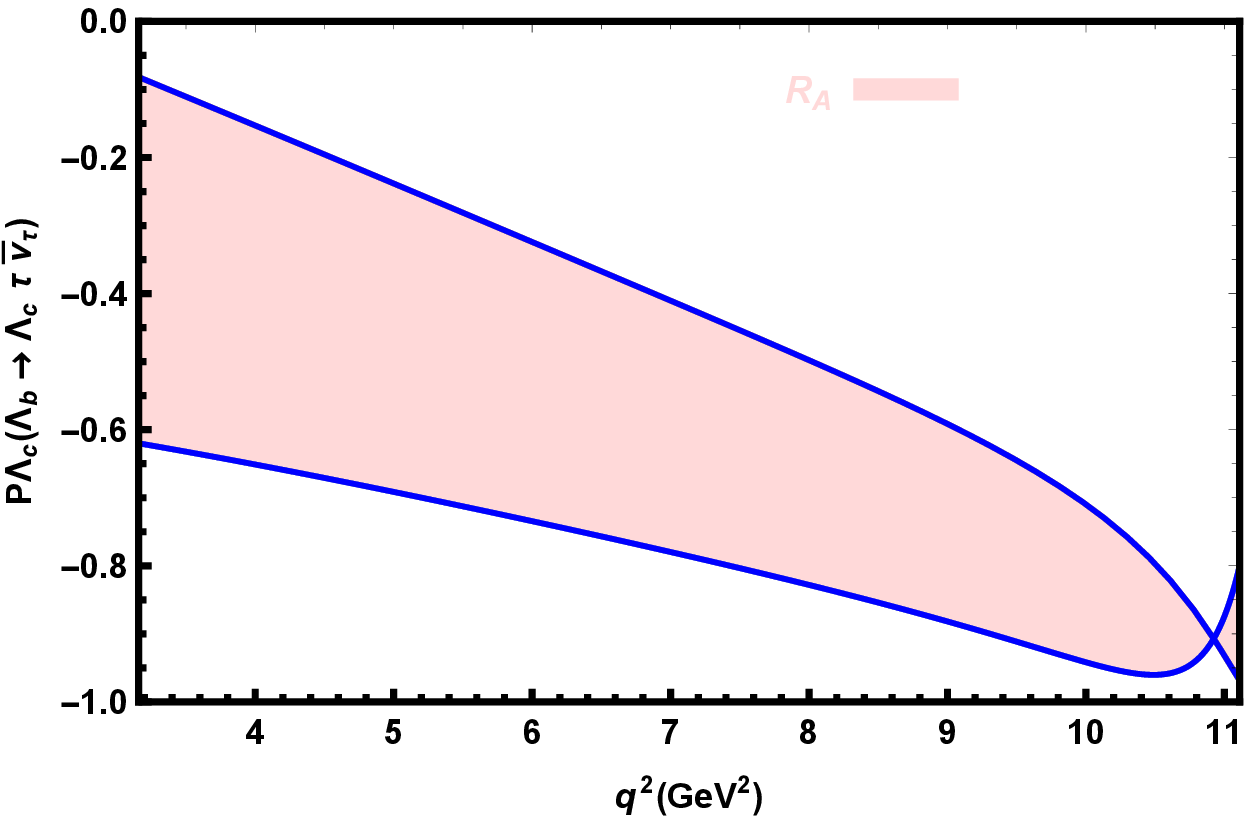}
\includegraphics[totalheight=6cm,width=8cm]{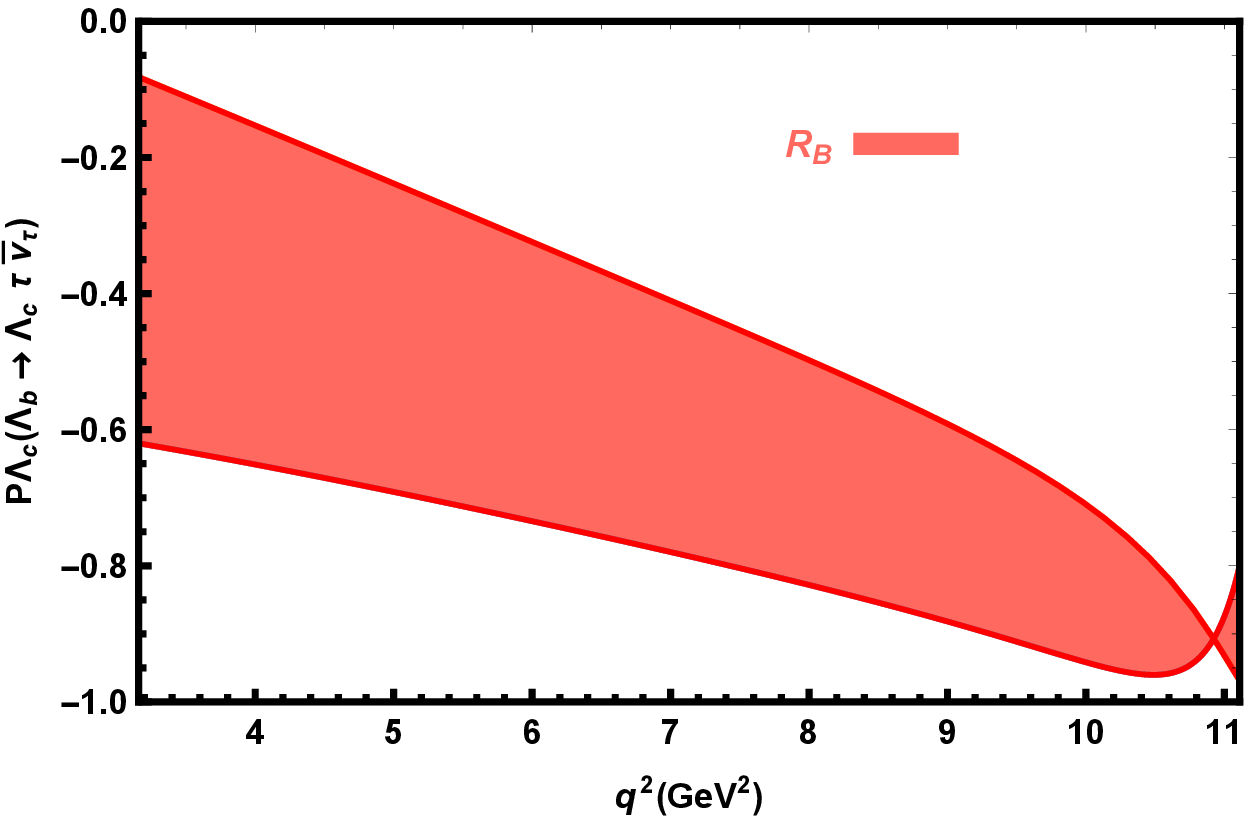}
\end{center}
\end{figure}
\begin{figure}[h!]
\begin{center}
\includegraphics[totalheight=6cm,width=8cm]{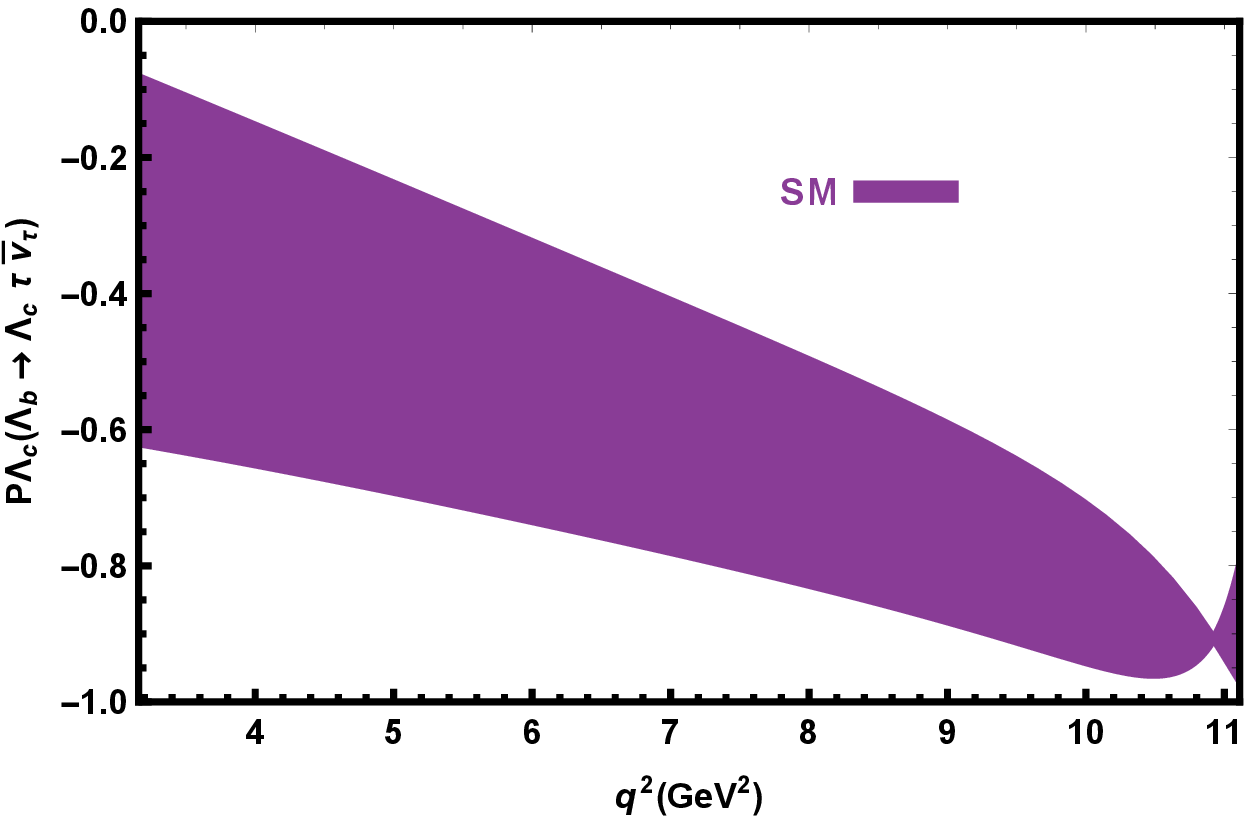}
\includegraphics[totalheight=6cm,width=8cm]{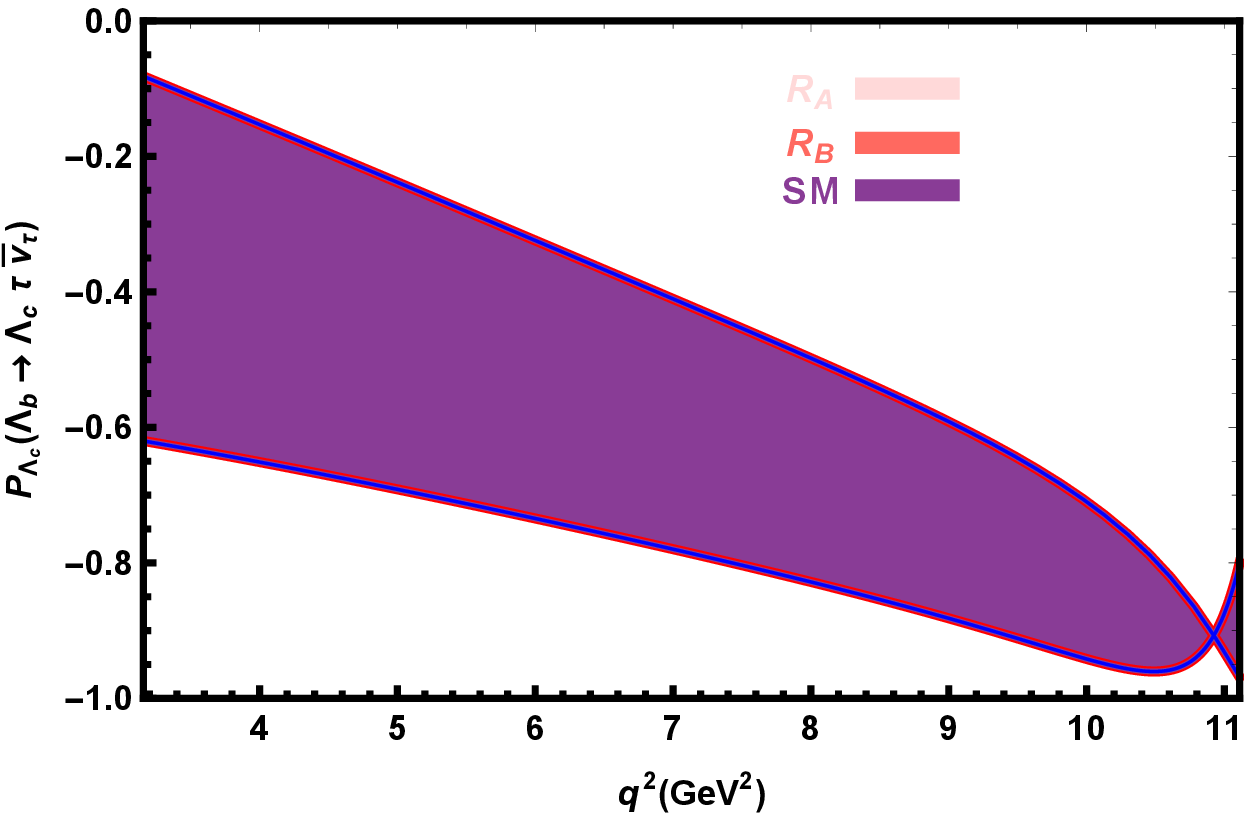}
\end{center}
\caption{The dependence of the $ P_{\Lambda_{c}} $ on  $q^2$  for the $\Lambda_{b}\rightarrow \Lambda_{c} \tau  \overline{\nu}_\tau$  transition   in  SM and vector LQ models with all errors.}
\end{figure}

\end{widetext}

\begin{figure}[h!]
\centering
\begin{tabular}{ccc}
\epsfig{file=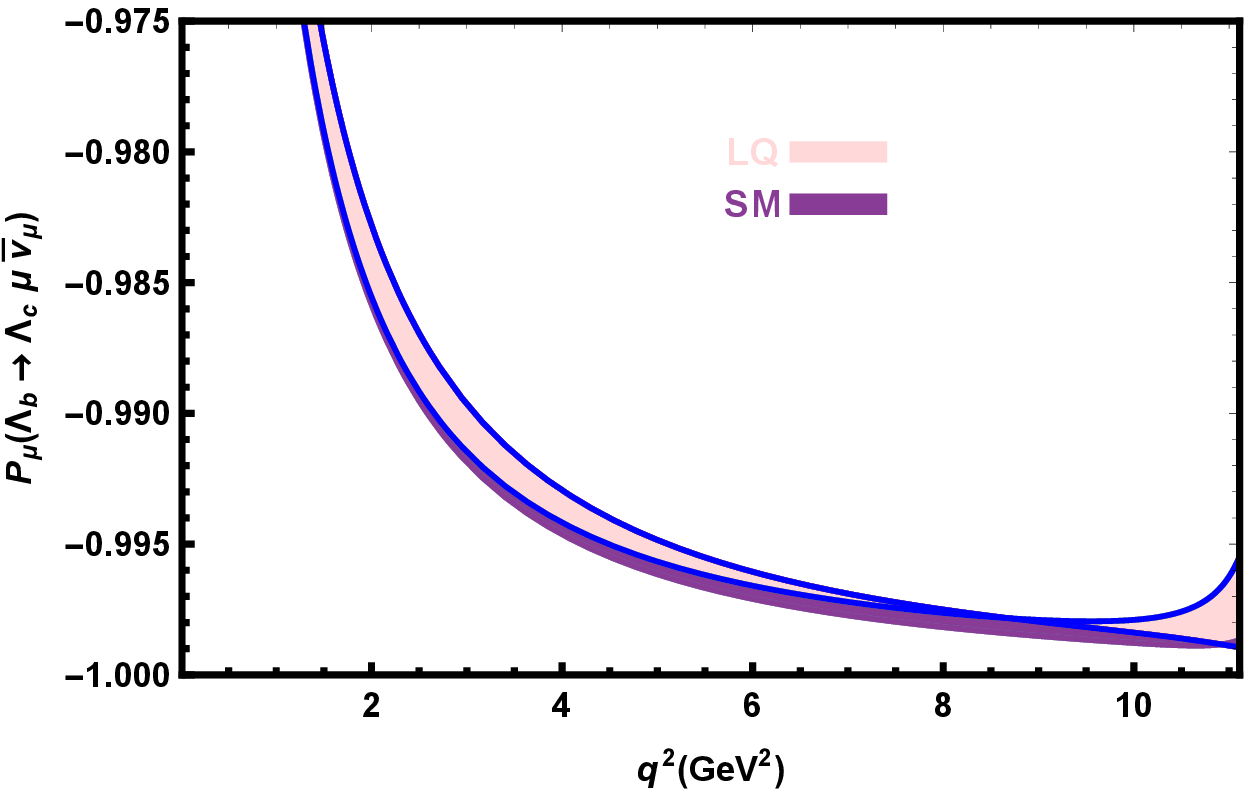,width=1.0\linewidth,clip=} 
\end{tabular}
\caption{The dependence of the $ P_{\mu} $ on  $q^2$  for the $\Lambda_{b}\rightarrow \Lambda_{c} \mu  \overline{\nu}_\mu$  transition   in  SM and vector LQ models with all errors.}
\end{figure}

\begin{widetext}

\begin{figure}[h!]
\begin{center}
\includegraphics[totalheight=6cm,width=8cm]{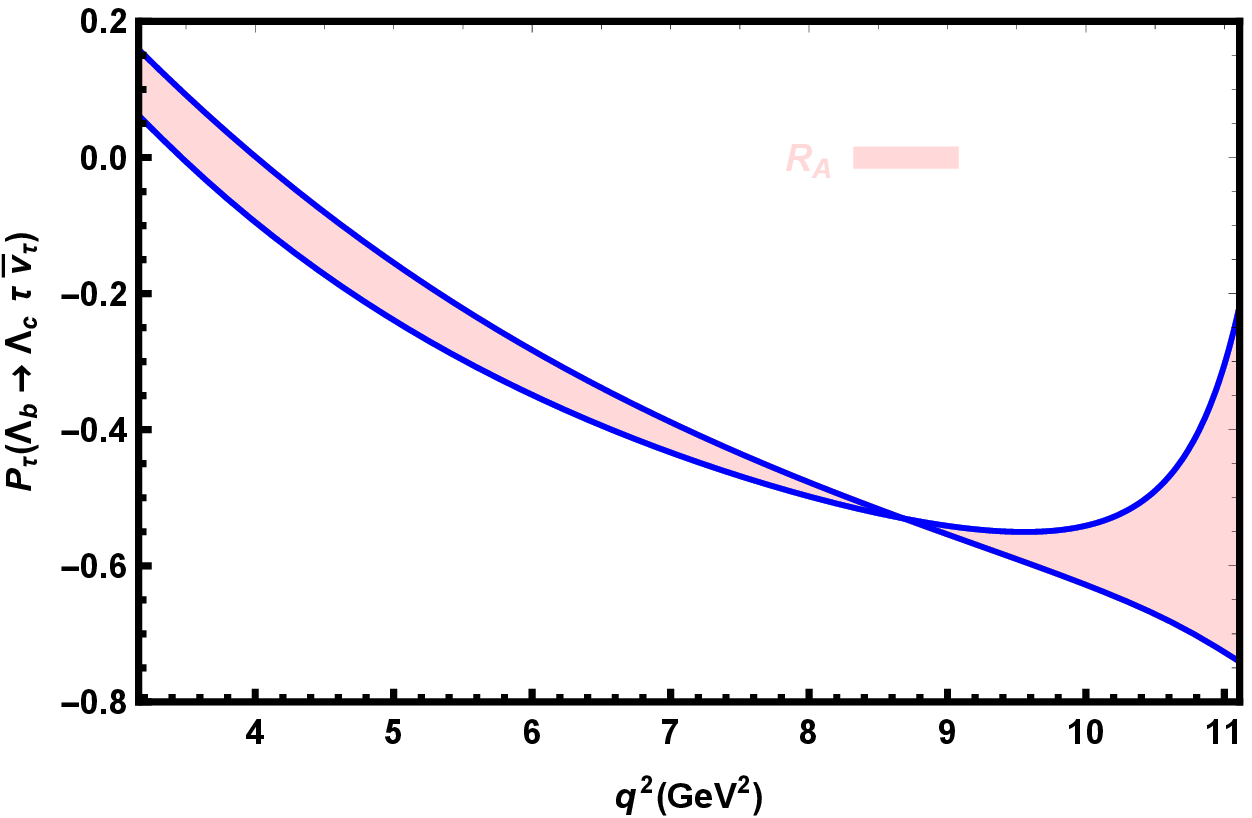}
\includegraphics[totalheight=6cm,width=8cm]{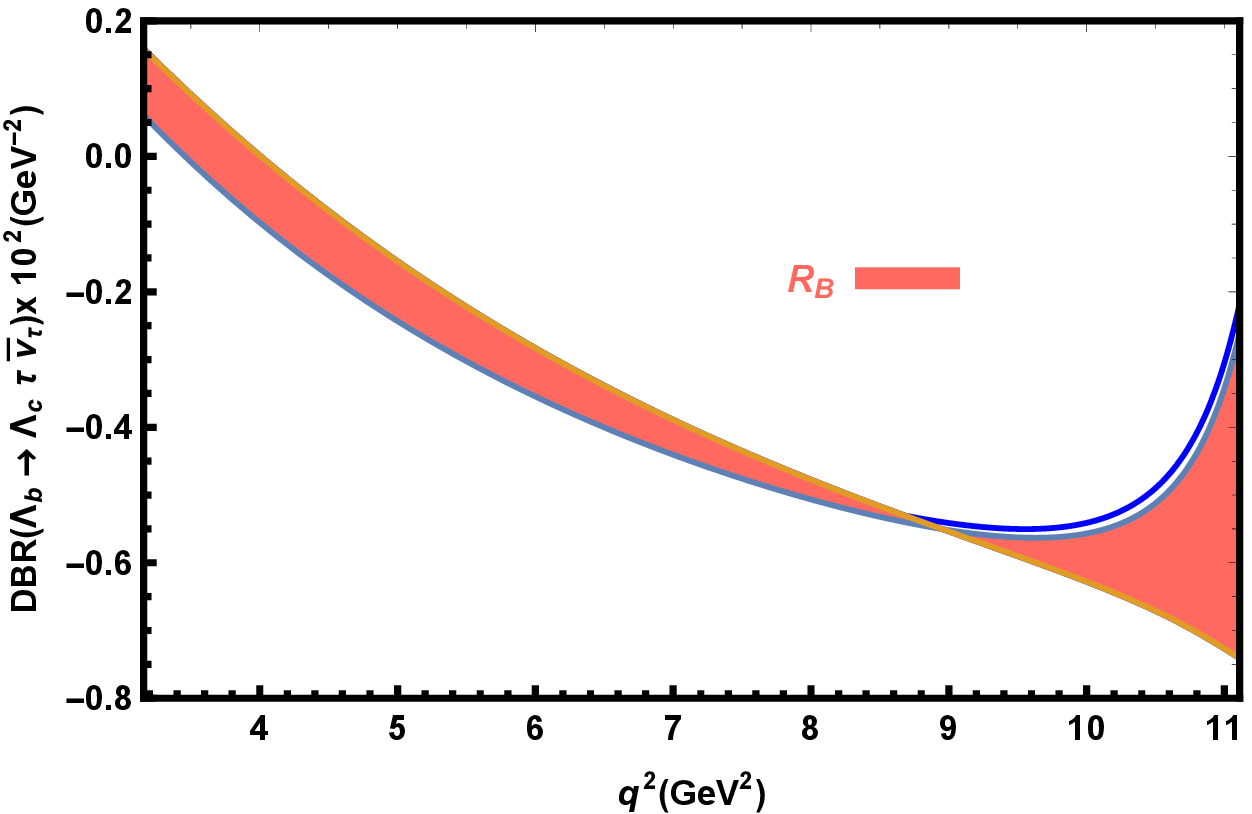}
\end{center}
\end{figure}
\begin{figure}[h!]
\begin{center}
\includegraphics[totalheight=6cm,width=8cm]{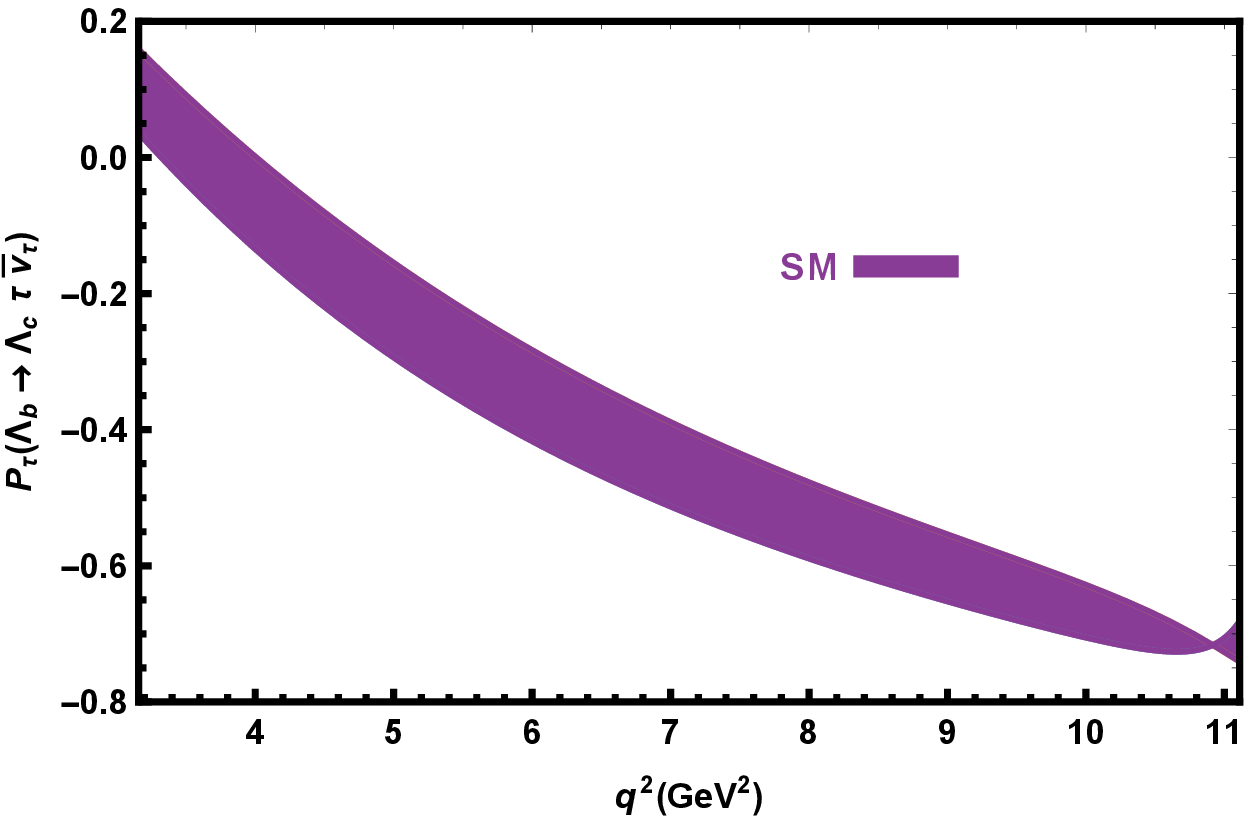}
\includegraphics[totalheight=6cm,width=8cm]{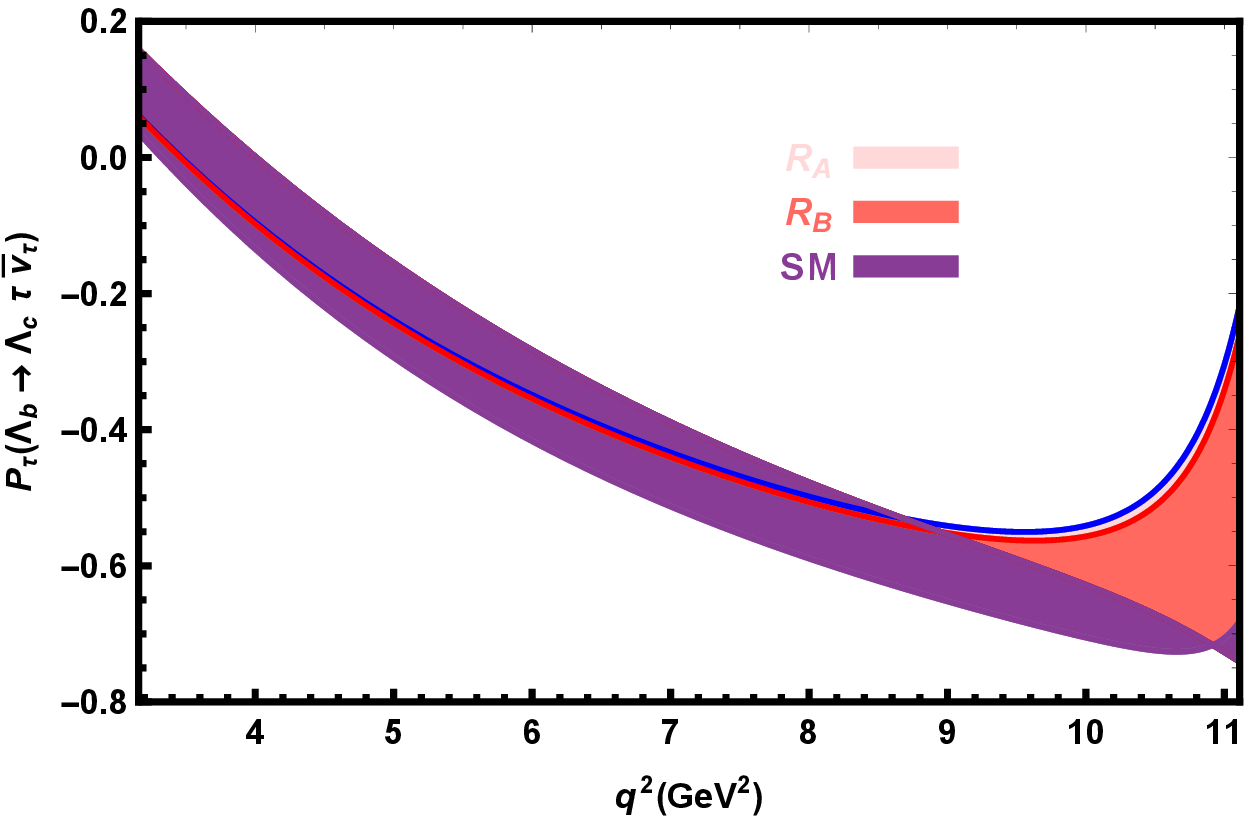}
\end{center}
\caption{The dependence of the $ P_{\tau} $ on  $q^2$  for the $\Lambda_{b}\rightarrow \Lambda_{c} \tau  \overline{\nu}_\tau$  transition   in  SM and vector LQ models with all errors.}
\end{figure}

\end{widetext}


\section{Summary and Conclusions}
The direct search for NP effects has ended up in null results so far.  There is a hope to hunt these effects indirectly in some hadronic decay channels. The experimental data on   $ R(D^{(*)}) $, $ R(K^{(*)}) $ and $ R(J/\psi) $ have shown sizable deviations from the SM predictions, recently. The test of similar possible deviations in baryonic sector is  of great importance. Different experiments may put in their agenda to answer  this question in near future. In this situation, theoretical and phenomenological studies can play important roles before the experimental results.  The anomalies between the data and SM predictions in the mentioned mesonic channels can be removed by introducing some NP scenarios BSM. Among these models are vector and scalar leptoquark models. We have  investigated the tree-level $  \Lambda_{b}  \rightarrow {\Lambda}_{c}  \ell \overline{\nu}_{\ell} $  in the SM and vector leptoquark models and compared the results with each other. Our aim is to provide results from different models, which may be compared with future experimental data. 

In particular we calculated the (differential) branching ratios and forward-backward asymmetries at $ \mu $ and $ \tau $ lepton channels and saw no deviations of the LQ results from the SM predictions and existing experimental data in $ \mu $ channel. In the calculations we used the form factors calculated in full QCD as the main inputs and took into account all the errors coming from the form factors and model parameters. At $ \tau $  channel, the results of both models on $\textit{A}_{FB}$ are the same, as well. This is an expected result since in the LQ model the NP effects are encountered via Wilson coefficients that appear in both the nominator and denominator in the $\textit{A}_{FB}$  formula and their effects are canceled.   However,  it is observed that at $ \tau $ channel,    the leptoquark models, especially the the $ R_A $ type fit solution,  sweep some regions out of the SM band on $ DBR(q^2)-q^2$ graph.

We also investigated the behavior of $R(q^2)$ with respect to  $q^2$ and extracted the values of the  parameter  $R(\Lambda_c)$ at different scenarios. We observed that the LQ predictions on $R(q^2)-q^2$ and  $R(\Lambda_c)$  using both  the $ R_A $ and  $ R_B$ type fit solutions deviate considerably from the SM predictions. 

Finally, we considered the the $\Lambda_{c}$ baryon and lepton  polarizations, which are important parameters to search for  new physics effects, as well. we observed  that the LQ and SM predictions show considerable differences on $ P_{\Lambda_{c}} -q^2$ in $ \mu $ channel. However, in $ \tau $ channel, the SM  and both the LQ scenarios have  roughly the same predictions on $ P_{\Lambda_{c}}-q^2 $. In the case of  lepton polarization, $ P_{\mu} $, we see small shifts in some regions between the SM and LQ predictions. As far as the $ P_{\tau} $ is concerned, the $ R_A $ and  $ R_B $ type LQ scenarios have almost the same predictions, but their results indicate considerable deviations from the SM prediction.

The overall considerable differences between the LQ and SM predictions on the parameters related to the  tree-level $  \Lambda_{b}  \rightarrow {\Lambda}_{c}  \ell \overline{\nu}_{\ell} $ transition detected in the present study make this mode as one of the  important  baryonic $ b\rightarrow c $ based  transition,  which may be considered as a good probe to search for NP effects. Future data on the physical quantities considered in the present study, which would be available after measurements on  $ \Lambda_b \rightarrow \Lambda_c \tau ~ \overline{\nu}_\tau$ channel, will be very useful in this regard. 

\label{sec:Num}

\end{document}